\documentclass[preprints,article,accept,pdftex,oneauthor]{Definitions/mdpi} 
\firstpage{1} 
\pubvolume{1}
\issuenum{1}
\articlenumber{0}
\pubyear{2023}
\copyrightyear{2023}
\externaleditor{Academic Editor: Peter Roming, Michael Siegel 
}
\datereceived{4 January 2023} 
\daterevised{15 February 2023} 
\dateaccepted{16 February 2023} 
\datepublished{} 
\hreflink{https://doi.org/} 
\pdfoutput=1

\newcommand\swift{{\it Swift}} 
\newcommand\swiftb{{\it Swift}/BAT} 
\newcommand\swiftx{{\it Swift}/XRT} 
\newcommand\swiftu{{\it Swift}/UVOT} 
\newcommand\tninety{${\rm T_{90}}$} 

\usepackage{caption}
\usepackage{subcaption}

\Title{Swift/UVOT: 18 Years of Long GRB Discoveries and Advances}

\TitleCitation{Swift/UVOT: 18 Years of Long GRB Discoveries and Advances}

\Author{Sam Oates 
 \orcidA{}}

\AuthorNames{Sam Oates}

\AuthorCitation{Oates, S.}

\address[1]{School of Physics and Astronomy \& Institute for Gravitational Wave Astronomy, University of Birmingham, Birmingham 
 B15~2TT, UK; s.r.oates@bham.ac.uk or samantha.oates@ucl.ac.uk
}

\abstract{The Neil Gehrels Swift Observatory (\swift) has been in operation for 18 years. The Ultra-Violet/Optical Telescope (UVOT) onboard 
\swift~was designed to capture the earliest optical/UV emission from gamma-ray bursts (GRBs), spanning the first few minutes to days after the prompt 
gamma-ray emission. In this article, I provide an overview of the long GRBs (whose prompt gamma-ray duration is >2 s) observed by the \swiftu,~and review 
the major discoveries that have been achieved by the \swiftu~over the last 18 years. I discuss where improvements have been made to our knowledge 
and understanding of the optical/UV emission, particularly the early optical/UV afterglow.}


\begin{document}

\section{Introduction}
Gamma-ray bursts (GRBs) are brief, intense flashes of gamma-rays, often accompanied by a longer-lasting emission in 
the X-ray to radio wavelengths. The duration of the gamma-ray emission may be as short as a few milliseconds or may last 
for as long as several hundreds of seconds, during which the GRB `outshines' all objects in the known universe. The initial 
gamma-ray emission is followed by an afterglow, longer-lived emission that is usually observed from X-ray through to radio wavelengths 
\citep{sari97,sar98}. 

GRBs can be divided approximately into two classes, short and long, by the duration of their gamma-ray emission, with the division at $\sim$2 s \citep{kou93}. 
The short GRBs are thought to be a result of the merger of two compact objects, either two neutron stars or a neutron star and a black hole \citep{pacz86,eic89}. This 
was recently confirmed (in one case) with the detection of a GRB associated with the GW signal of binary neutron star merger (GW 170817/AT2017gfo; e.g.,~\citep[]{abb17,abb17b,cou17,eva17,gol17}). Long GRBs are thought to be the collapse of massive stars \citep{woo93, mac99}; evidence for this is the 
association of supernovae a few days after the gamma-ray emission (e.g.,~\citep{gal98,kul98,hjo03,woo06}).

Following the launch of the Neil Gehrels Swift Observatory ({henceforth, \it Swift}; \citep[]{geh04}), it was shown that the empirical categorisation of 
short and long GRBs, by their gamma-ray duration, quantified by the \tninety~parameter (${\rm T_{90}}$ is the time interval over which 90\% of the gamma-ray emission is measured) with a division at 2 s, is too simplistic. There is likely some natural overlap in duration between the classes \citep{bro13}. Some GRBs that are categorised as long or short have been shown to have other properties that suggest that they belong to the other class. The prototypical event was the {\it Swift} discovered GRB 060604 \citep{del06,fyn06,geh06,gal06}. More recent examples of each type are GRB 200826A \citep{ahu21,zha21,ros22} and GRB 211211A \citep{ras22,tro22,yan22}. In addition, different detectors with different sensitivities and different methods for determining $\rm T_{90}$ may result in slightly different values for $\rm T_{90}$, adding to the complication of classifying GRBs solely by their duration (e.g., \citep[]{sak11b,sav12,bro13,qui13}). Because of such complications, there is some preference to categorise GRBs by their progenitors, into merger and collapsar categories, which avoids the pitfalls of classifying GRBs by their gamma-ray durations. However, classifying GRBs into merger (Type I) and collapsar (Type II) classes \citep{zha06b} does often require a more involved analysis and may not be possible for all observed GRBs, for instance, due to insufficient observations. For simplicity, in this review, I will use the empirical short and long GRB dichotomy unless otherwise stated. 

In this review, I will discuss the advances in long GRBs and their afterglows as a result of optical/UV observations performed by the Ultra-violet/optical telescope (UVOT) onboard \swift. The review is divided into Section \ref{preSwift}, a discussion of the discovery and status of long GRBs before the launch of {\it Swift}; Section \ref{UVOTera}, an overview of long GRBs observed with \swiftu; Section \ref{UVOTdiscoveries}, advances in our understanding of the optical/UV emission as a result of \swiftu;~and summarise in Section \ref{conclusions}. A review of short GRBs observed by \swiftu~is provided in an accompanying article in this special issue.

\section{Long GRBs {Pre-{\it Swift}}}
\label{preSwift}
In this section, to put \swiftu~and its discoveries into context and where it fits into the timeline of long GRB discoveries, I provide a short overview 
of our understanding of long GRBs up until the launch of {\it Swift}.

\subsection{The Discovery of GRBs}
Gamma-ray bursts were initially detected during the late 1960s by the 4 Vela spacecraft~\citep{kle73}. These unknown sources of gamma-rays were found 
not to be associated with any object within our solar system, and therefore were assumed to be of Galactic or extra-galactic origin. Further progress 
had to wait until the launch of the Compton Gamma Ray Observatory (CGRO) in 1992, where, using the Burst Alert Transient Source Experiment 
(BATSE) onboard the CGRO, it was found that the distribution of GRBs was isotropic, suggesting a cosmological origin \cite{meg92}. Also discovered 
using BATSE was the bimodal distribution of the duration of the gamma-ray emission: GRBs with hard spectra and of $<$2~s duration were 
classified as short-hard GRBs, and those with soft spectra and of $>$2~s duration were classified as long-soft GRBs \cite{kou93}, providing the first 
evidence for two different types of progenitors. 

Emission at longer wavelengths, X-ray to radio, was predicted. This emission, the `afterglow', was expected to be longer lived and to be observed shortly after the prompt 
gamma-ray emission \citep{liang85,hart88,rap85}. A detection was only achieved at longer wavelengths, with the launch of BeppoSax in 1996, which 
had instruments to observe 0.1--300~keV X-rays~\mbox{\citep{boe97,par97,boe97b,fon97}}. The first GRB to be detected in X-rays, and subsequently in the optical, was GRB~970228 \citep{cos97,van97}. Observations of GRB~970508 provided the first redshift, $0.835\leq z\leq2.3$ \citep{metz97}, confirming the extra-galactic origin of these sources, and provided the first indications of how energetic these sources are, with an isotropic energy of $7\times 10^{51}$ erg calculated for this GRB \citep{metz97}. One last further major development was the detection of supernovae coincident with long GRBs; the first was SN1998bw, associated with GRB 980425 \citep{pia99,gal98}; and the second, SN2003dhm associated with GRB 030329 \citep{hjo03}. With GRB 980425, there was some concern that the supernova was a chance coincidence, but with a second GRB associated with a supernova, the connection between the collapse of massive stars and long GRBs was solidified \citep{woo06}.  

\subsection{Expectations and Observations of Long GRBs, Pre-Swift}
\label{expectations}
Prior to the launch of \swift, the theoretical understanding of long GRBs and their emission in the optical/IR was well developed. The collapse of the 
massive star to a black hole was predicted to result in the production of relativistic bipolar jets \citep{kat97}. The relativistic jets can be regarded 
as a series of relativistic shells with varying Lorentz factors \citep{kob97, kob02}, which result in internal shocks when the shells with high Lorentz factors 
catch up with shells of lower Lorentz factor that were emitted earlier. The internal shocks produce the prompt gamma-ray emission. Once the relativistic 
jet reaches the external medium and ploughs into it, it begins to slow down and produces the external shocks, which comprise of a forward and a reverse shock \citep{sari97}. The forward shock propagates into the external medium, while the reverse shock travels back through the relativistic ejecta. The emission produced in both 
shocks is released by synchrotron radiation. This emission results in a long-lived afterglow, which for the most part is emission from the forward shock. The 
emission produced by the external shocks can be described by power-laws for both the observed spectra and the light curves. The temporal and spectral indices, 
$\alpha$ and $\beta$, are given using the standard convention of $F(t)\propto t^{\alpha}\nu^{\beta}$, where $t$ is time and $\nu$ is frequency. The expected 
value of the spectral and temporal indices can be described analytically by a set of closure relations~\citep{sar98}. For a given observing band, the choice of 
closure relation depends on its location relative to the synchrotron frequencies (the synchrotron cooling frequency $\nu_c$, the synchrotron peak frequency $\nu_m$, and the synchrotron 
self-absorption frequency $\nu_a$), which in turn depends on the values of the microphysical parameters (the kinetic energy of the outflow $E_k$, the fraction 
of energy given to the electrons $\epsilon_e$, the fraction of energy given to the magnetic field $\epsilon_B$, the structure and density of the external medium,
and the electron energy index $p$). 

The behaviour of the afterglow light curve is also dependent on whether or not the observer is seeing all of the emission from the jet, which has 
an opening angle $\theta$. Initially, the radiation contained within the jet will move at a bulk Lorentz factor $\Gamma$ and will be beamed into an 
opening angle $\Gamma^{-1}$, which is smaller than the jet opening angle, $\theta$. As $\Gamma$ decreases, the observer will begin to see more and 
more of the jet until they see the emission from the entire jet at $\Gamma\sim\theta^{-1}$ \citep{rho97,piran99}. At this point, an achromatic break in the light curve will occur, known as the `jet break', after which the light curves will decay at a much steeper rate~\citep{rho97,rho99,sari99}. 

Prior to the launch of \swift, the number of GRBs with optical/IR detections was small, around 60 GRBs had optical/IR detections, and only a handful of these 
had well-sampled optical/IR light curves \citep{zeh06}. The optical/UV behaviour was consistent with power-law behaviour, often with a break in the light curve. The average power-law 
indices of optical/IR afterglows with observed breaks in their light curves were $\alpha_{1}=1.05\pm0.1$ and $\alpha_{2}=2.12\pm0.14$ \citep{zeh06}. The first decay 
index is consistent with the predictions of the closure relations for an isotropic outflow, while the latter decay index is consistent with the decay expected 
after a jet break. Observations of the optical emission typically began a significant amount of time after the gamma-ray detection, commencing typically 
more than 0.1 days after the trigger \citep{zeh06}. This left a significant gap in our knowledge of the early optical/IR emission, between the onset of the 
GRB through to the first few thousand seconds after the trigger. There were also questions as to why not all GRBs had bright optical or radio afterglows, 
while X-ray afterglows were expected for all GRBs \citep{geh04}. It was thought that these GRBs were `dark', potentially due either to dust extinction \citep{lam03}, 
being (intrinsically) faint and thus not detected when observations occurred several hours after the trigger~\citep{cre03, pan03b}, that the optical 
decayed more rapidly 
than the X-ray \citep{gro98a}, or that they were at high redshift \citep{geh04}. Prompt high-quality X-ray, UV, and optical observations over the first minutes to hours 
of the afterglow were thus crucial to obtain a better understanding of~GRBs.

\section{The Era of  \emph{Swift}/UVOT}
\label{UVOTera}
\swift, launched in November 2004, has been in successful operation for 18 years. \swift~was designed 
specifically to detect GRBs, and observe rapidly with its narrow field instruments. \swift~houses three instruments that are designed to 
capture gamma-ray, X-ray, and optical/UV emissions. The Burst Alert Telescope (BAT \citep[]{bar05}) detects the prompt gamma-ray emission, while the 
narrow field instruments, the X-ray Telescope (XRT \citep[]{bur05}), and the Ultra-violet/Optical Telescope (UVOT \citep[]{roming}), observe the X-ray 
and optical/UV afterglow, respectively. The co-alignment of the XRT and \swiftu~instruments is ideal for observing GRB afterglows because it allows the X-ray and 
optical/UV emission to be observed simultaneously. \swift's capacity to slew rapidly to point the narrow field of view telescopes at the GRB location, enables regular and 
early observations of GRBs at longer wavelengths, starting as early as a few tens of seconds after a BAT trigger.

\subsection{An Overview of Long GRBs Observed with \swiftu~}
\label{uvot_overview}
At the time of writing \swiftu~~has observed $\sim$1339 long GRBs (\url{https://swift.gsfc.nasa.gov/archive/grb\_table/}; accessed on 20th October 2022) with $\rm T_{90}>2$\,s. The majority of GRBs observed by \swiftu~were initially detected by \swiftb, resulting in \swift~automatically slewing to point the \swiftx~and \swiftu~instruments at the GRB location to rapidly commence observations in the X-ray and optical/UV. However, for a small number of GRBs, \swiftu~began observations much later, since the gamma-ray emission did not trigger \swiftb, but was detected either through BAT ground analysis, or by other X-ray/gamma-ray observatories, including the High Energy Transient Explorer-2 (HETE2), the INTErnational Gamma-Ray Astrophysics Laboratory (INTEGRAL), Konus-Wind, the Interplanetary Network (IPN), Fermi, Astro-rivelatore Gamma a Immagini Leggero (AGILE), and MAXI and CALET.

After an automatic slew, \swiftu~takes its first exposure. This short, typically 10 s, $v$ band exposure is taken while the reaction wheels are stabilising \swift, such that the pointing direction  is in transition from a motion of many arcseconds per second to a stable pointing. During the slew, \swiftu~is protected from damage by bright stars passing through the field of view by maintaining the photocathode at a low-voltage state. As such, the settling exposure is frequently ignored or discarded \citep{oates12}, since the photometry may be uncertain due to the \swiftu~photocathode voltage changing at the beginning of the exposure, or concerns regarding the rapidly changing spacecraft attitude \citep{rom17}. Ref.~\cite{pag19} examined the settling exposure for 26 GRBs detected by \swiftu. They found an issue in only 2 out of the 26 GRBs, and only in the first second of the settling exposures. Fortunately, the settling exposure is taken in event mode (in event mode, the arrival times and positions of the individual photons are recorded) and therefore, when issues arise, it does not have to be discarded completely, because it can be cut to exclude the affected parts of the exposure. The settling exposure usually begins 10--15 s before the first settled exposure. The fastest time to settled observations commencing for a long GRB is 52 s after the BAT trigger for GRB 050509A, and the average time to the start of settled observations is 120 s (calculated using GRBs observed by \swiftu~within the first 500 s, observed as a result of an automatic slew).

\begin{figure}[H]

\includegraphics[width=0.95\textwidth]{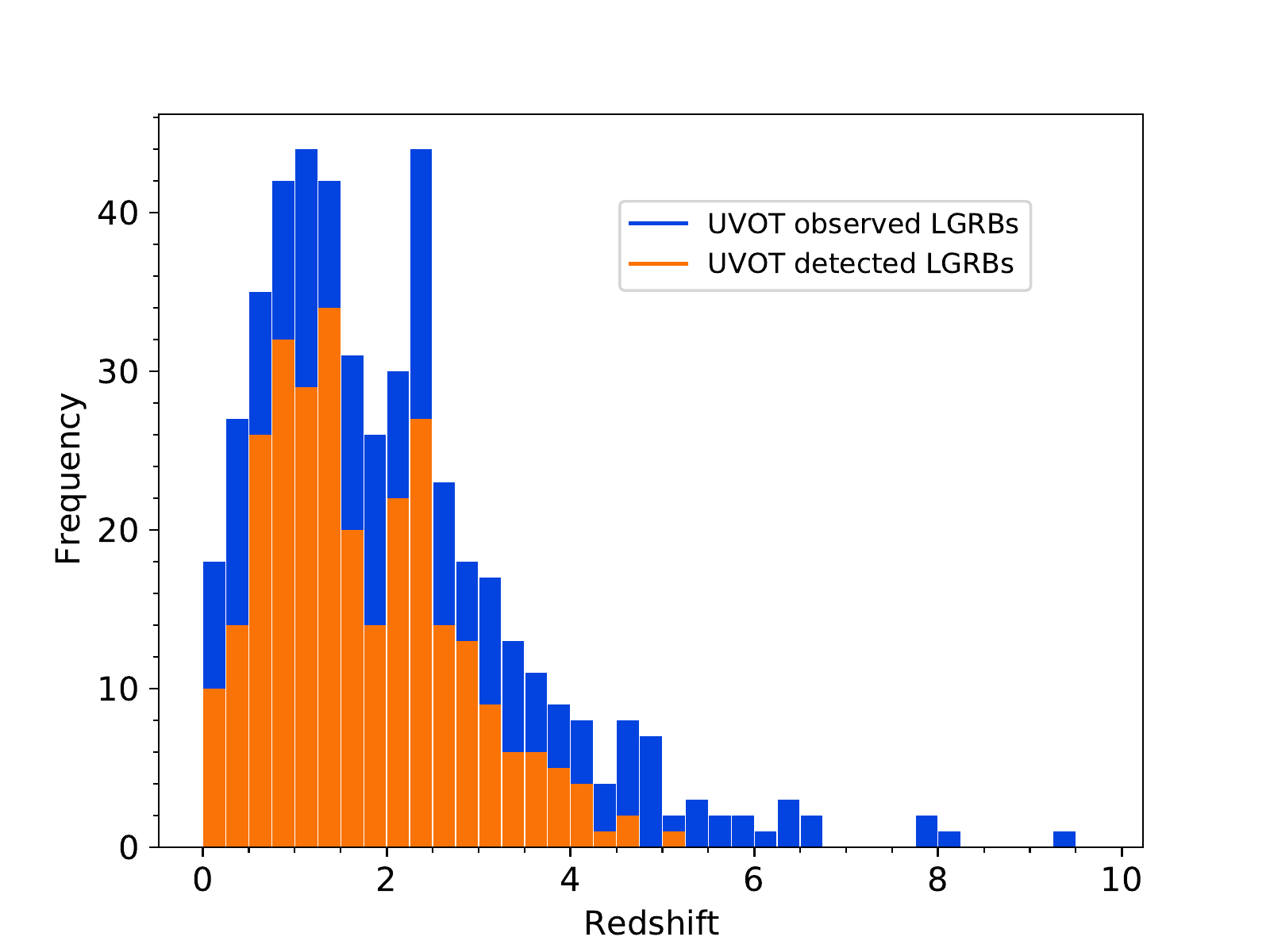}
\caption{The redshift distribution of \swiftu~GRBs. The redshift distribution of long GRBs observed by \swiftu~is given in blue, while the redshift distribution of long GRBs detected by \swiftu~is given in orange.}
\label{fig:redshift_distribution}
\end{figure}

\begin{figure}[H]
\includegraphics[width=0.9\textwidth]{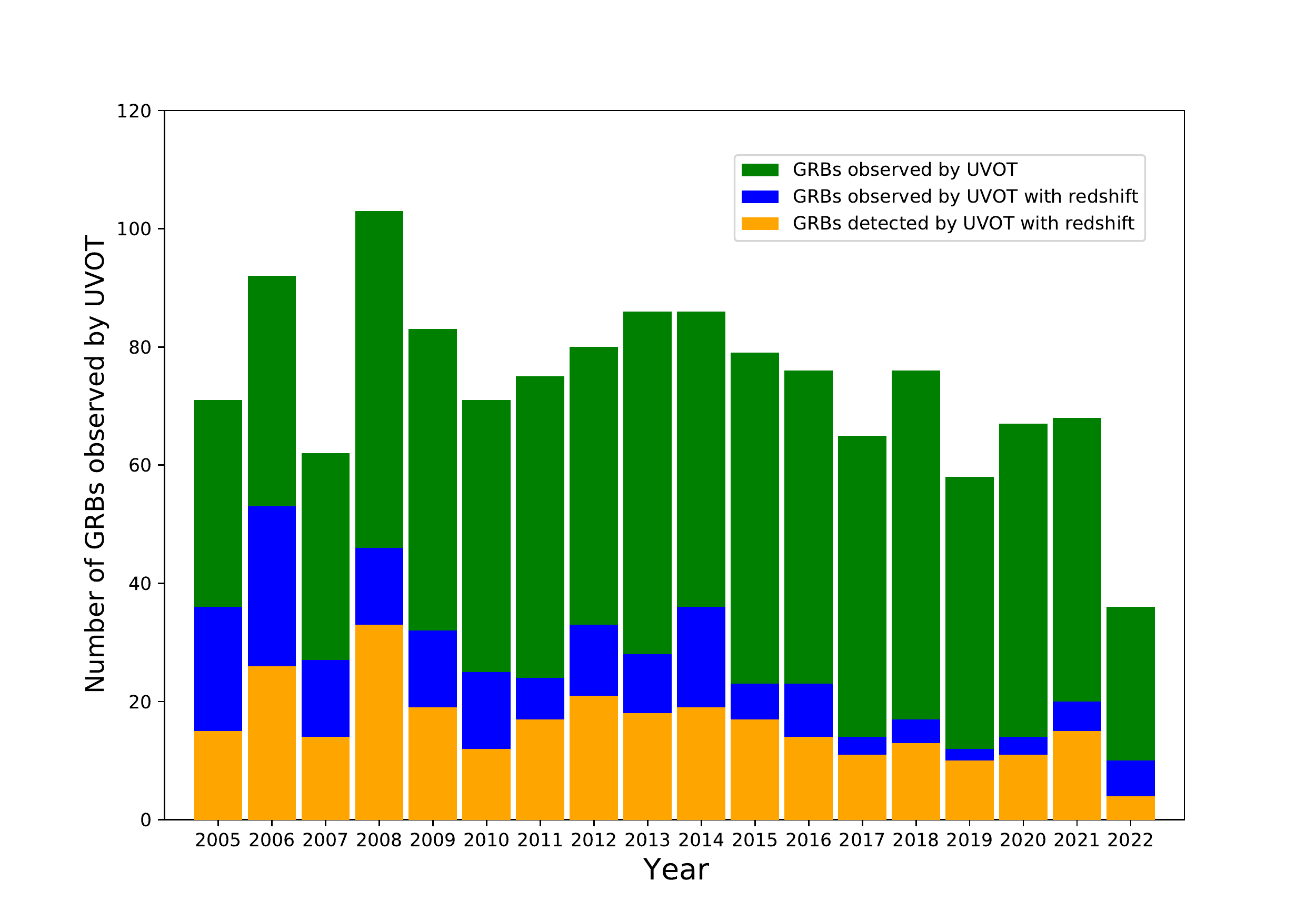}
\includegraphics[width=0.9\textwidth]{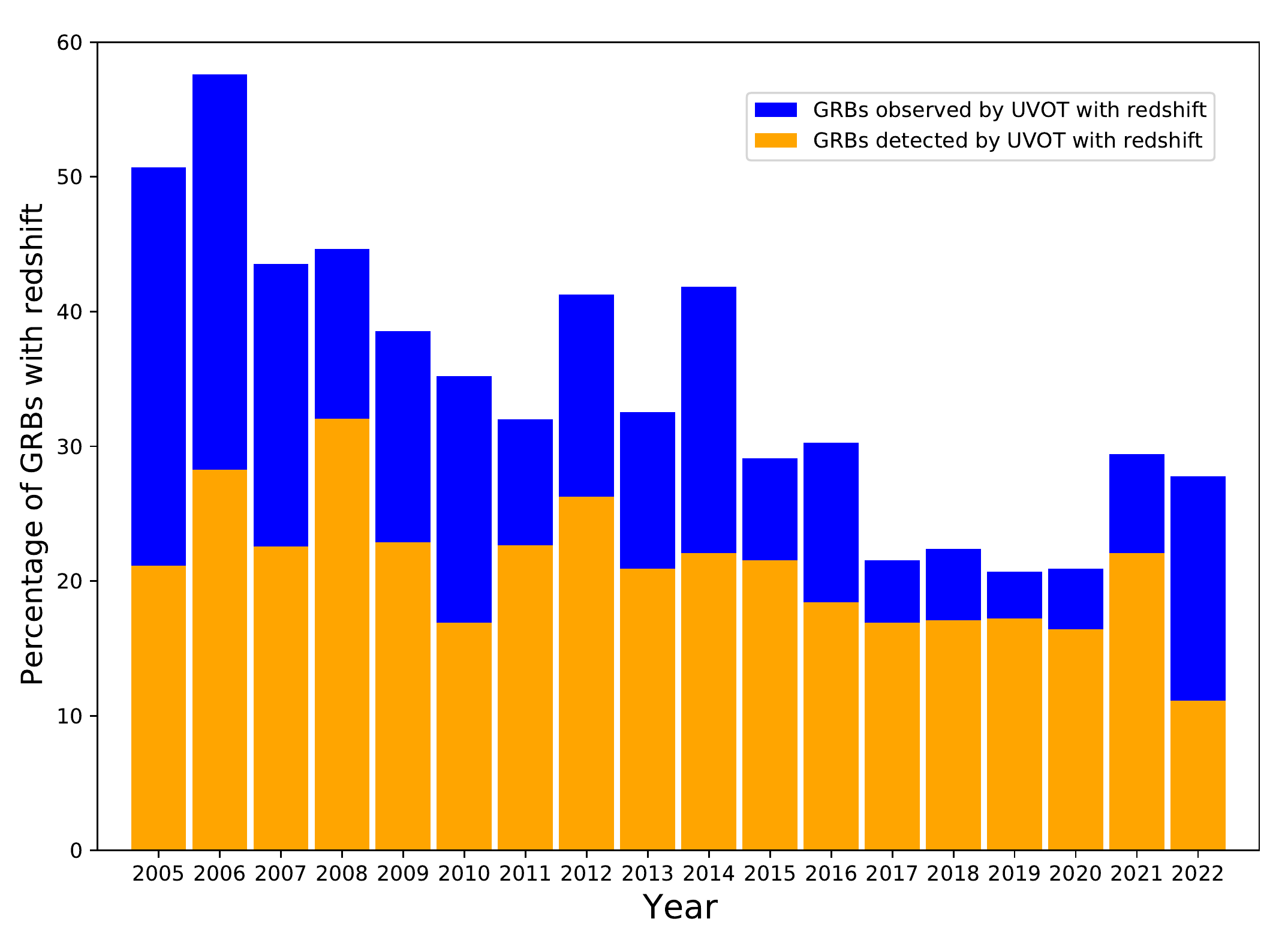}
\caption{The number of long GRBs observed each year by \swiftu. The top panel shows the number of GRBs, while the bottom panel shows the percentage. The green distribution provides the total number of long GRBs observed by \swiftu~per year. The blue distribution provides the total number/fraction of long GRBs observed by \swiftu,~which also have redshift, and the green distribution shows the total number/fraction of long GRBs detected by \swiftu,~which also have redshift. The fraction of long GRBs with redshift and observed by \swiftu~has steadily decreased with time; however, the fraction of long GRBs with redshift and that have a \swiftu~detection has not had such a significant reduction.}
\label{hist}
\end{figure}

Of the $\sim$1339 long GRBs observed by \swiftu, 514 have a reported \swiftu~detection in at least one \swiftu~filter (the \swiftu~filters are $white$, $v$, $b$, $u$, $uvw1$, $uvm2$, and $uvw2$ \citep{roming,poole,bre11}); 415 are within the first 500 s after the BAT trigger. This equates to a detection rate of 42\% if observed within the first 500 s. This value is consistent with that determined from the second \swiftu~catalogue, which, when using a sample of 538~GRBs, reported a detection rate of 43\% (with a detection threshold of at least $3 \sigma$) for long GRBs observed within the first 500 s \citep{rom17}. If considering a detection threshold of $2\sigma$, rather than $3\sigma$, the detection rate increases to 64\% \citep{rom17}. This is in contrast to 80--90\% of GRBs having optical/IR detections observed by ground-based telescopes \citep{cenko09,gre11}. Examining the number of GRBs with a photometric or spectroscopic redshift, 477 GRBs (36\%) observed by \swiftu~have redshifts (these GRBs may or may not have been detected by \swiftu); of these, 289 (22\%) have at least one detection in \swiftu. The redshift distribution for \swiftu~GRBs is given in Figure \ref{fig:redshift_distribution}. The average redshift for the \swiftu~observed GRBs is 2.0, while for the \swiftu~detected GRBs, the average redshift is 1.7. Figure \ref{hist} displays the total number of GRBs observed by \swiftu~per year, together with the number of GRBs observed by \swiftu~with a measured redshift per year, and the number detected by \swiftu~with a measured redshift per year. The second panel also provides the fraction of GRBs observed by \swiftu~with redshift and the fraction of GRBs detected by \swiftu~with redshift. It is concerning that the number of GRBs observed with \swiftu~with a reported redshift has decreased over the last 5--10 years, the fraction of GRBs observed with redshift in the first 5 full years of operation (2005--2009) is 47\%, while in the last 5 years of operation (2017--2021), this has reduced to 23\%. However, while the number of GRBs detected by \swiftu~with redshift shows a similar decreasing trend, the reduction is not as large, changing from 25\% in the first 5~years to 18\% in the last 5 years.

\begin{figure}[H]
\includegraphics[width=0.7\textwidth, trim={1.5cm 3.cm 1.5cm 12.5cm},clip]{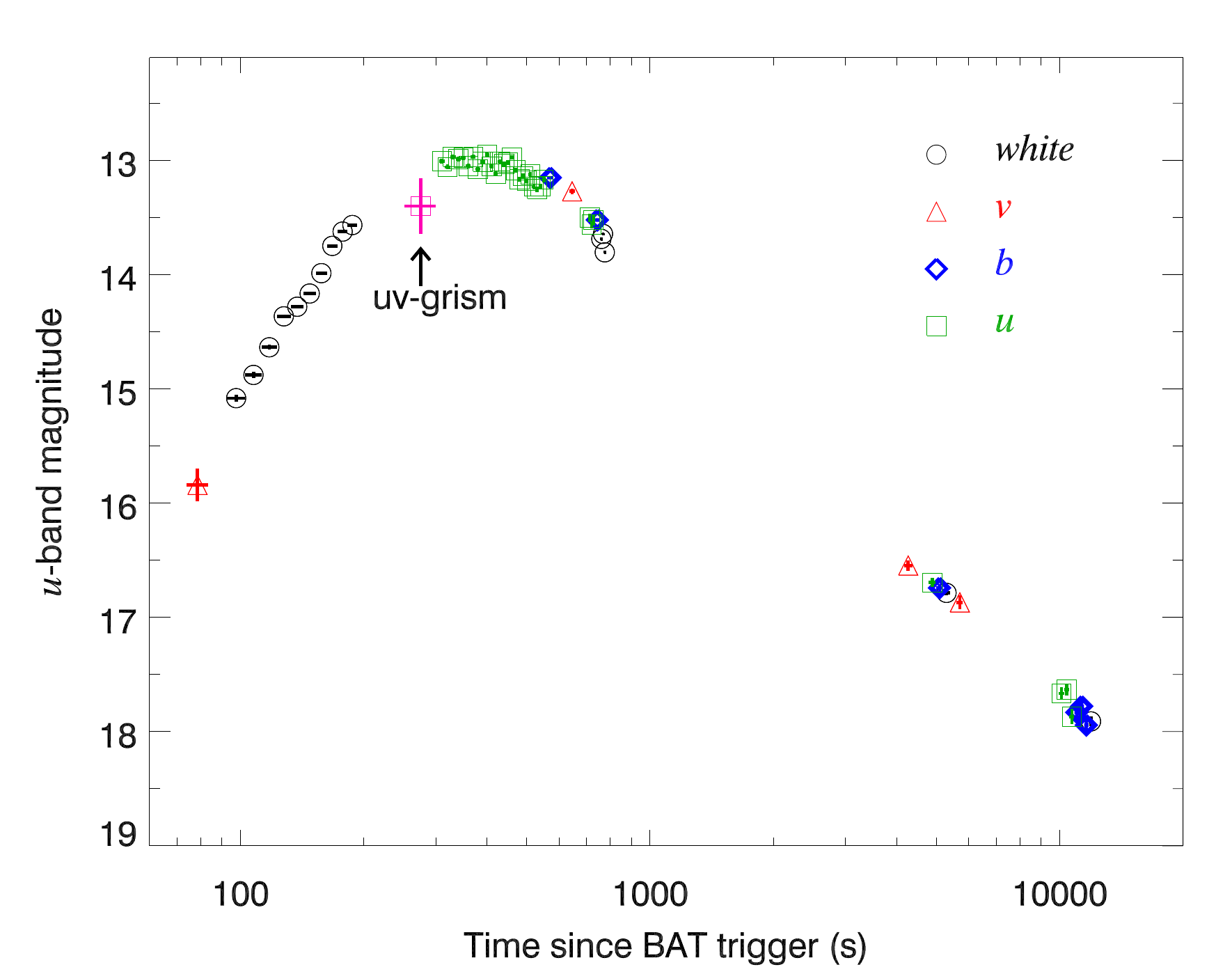}\\
\includegraphics[width=0.55\textwidth,angle=270, trim={1.5cm 0.cm 0.cm 3cm},clip]{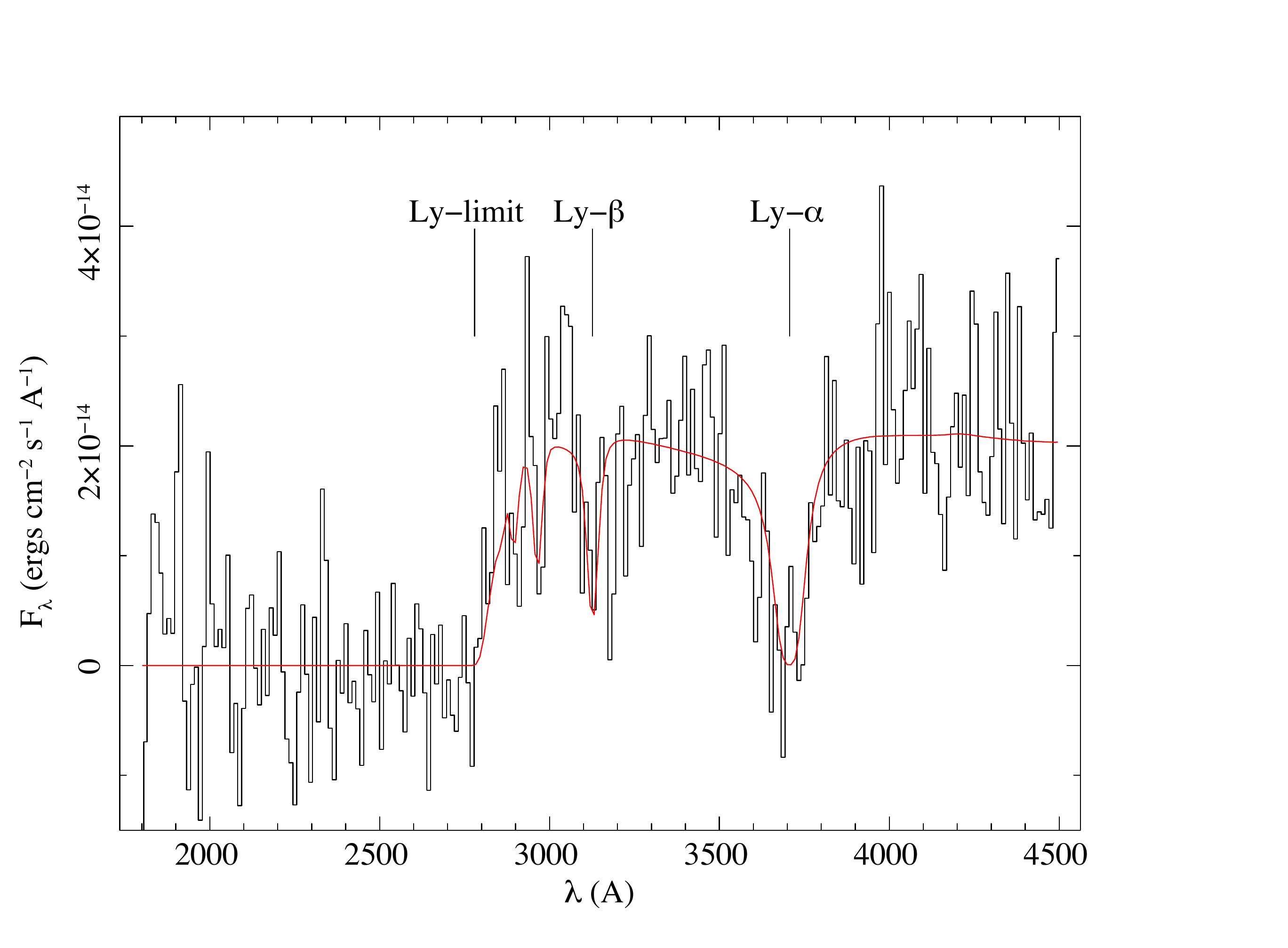}
\caption{\swiftu~observations of GRB 081203A (Figures are reproduced from \citep{kui09}, corresponding to Figures 1 and 3 in their article). The top panel 
displays the optical/UV \swiftu~light curve of GRB 081203A, including the photometric point derived from the \swiftu~UV-grism spectrum. The bottom panel displays 
the UV-grism spectrum of GRB 081203A. The red line is the best-fit model, which includes absorption by Lyman-$\alpha$, Lyman-$\beta$, and the Lyman forest. 
With these spectral features, a redshift of $2.05\pm0.01$ is derived for this GRB \citep{kui09}.}
\label{grism081203A}
\end{figure}

In addition to the six lenticular optical/UV filters, \swiftu's filter wheel also contains two grisms. These provide low-resolution field spectroscopy in the ultraviolet and
optical bands. The UV grism covers the range $\lambda$1700--$5000$\,\AA~ with a spectral resolution ($\lambda/\Delta\lambda$) of 75 at $\lambda 2600$\,\AA~ for the 
source magnitudes of $u$ = 10--16 mag, while the visible grism covers the range $\lambda$ 2850--6600\,\AA~ with a spectral resolution of 100 at $\lambda 4000$\,\AA~ for source 
magnitudes of $b$ = 12--17 mag \citep{kui15}. Since November 2008, the automated response sequence of the \swiftu, which governs the early exposures after a BAT
trigger \citep{roming}, includes a 50 s UV grism exposure, provided that the burst is bright enough in the gamma-rays. So far, this has resulted in two well-exposed 
UV spectra for GRB afterglows: GRB 081203A~\citep{kui09} and GRB 130427A \citep{mas14}. The light curve and grism spectrum for GRB 081203A is given in Figure~\ref{grism081203A}. \swiftu~has also been able to increase the number of GRBs with known redshift by obtaining photometric redshifts through the analysis of spectral energy distributions built using XRT and \swiftu~observations of the afterglow, supplemented by ground-based photometry (where available) \citep{depas07,kru11,oates12,gup21}.

\section{\emph{Swift}/UVOT~Long GRB Discoveries} 
\label{UVOTdiscoveries}
GRB 050318 was the very first GRB to be detected by \swiftu~\citep{sti05}. Observations began 3200 s after the BAT trigger, with an initial detection in the $v$ band of 17.8 mag. The optical emission was observed to decay, and the derived spectral and temporal indices were consistent with the expectations of the standard afterglow model, with the 
fireball expanding into a constant density medium, and the synchrotron cooling frequency lying in between the optical and X-ray bands. 

Since GRB 050318, we now have over 1300 GRBs observed with \swiftu,~and over 500 with a detection by \swiftu. Importantly, \swiftu~has provided a view of the 
first minutes to hours after the detection of the gamma-ray emission, which was lacking prior to \swift,~and which provided large numbers of GRBs with simultaneous 
X-ray and optical/UV observations. While the standard afterglow model, as described in Section \ref{expectations}, has been shown to be a good basic representation 
of the afterglow emission of \swift~GRBs, post-\swift~observations of GRBs often require a more complex model, for instance, with additional emission components 
or complex jet geometry, which I discuss further in this section. Many advances in GRB science have been through the analysis of multi-wavelength observations, 
and so, simultaneous observations with \swiftu~and \swiftx~have been key in unravelling the temporal and spectral behaviour of GRBs. In the rest of this review, 
I will focus on the most significant advances in GRB science as a result of \swiftu's 18 years of operation. I will discuss the most momentous GRBs observed, 
which due to their brightness, are the best-sampled and best-studied events. I will also discuss rare and usual GRBs observed by \swift, and improvements to our 
understanding of dark GRBs, the morphology of optical/UV light curves and correlations discovered with optical/UV~observations.

\subsection{Record-Breaking GRBs}
\label{record}
The detection of GRB 050318 was just the start of many major discoveries by \swift~and \swiftu. Over the years, \swift~has detected several GRBs that have pushed the 
boundaries of our expectations and knowledge. Some of the most striking and informative are the brightest GRBs observed via UVOT. These GRBs have the best signal-to-noise 
\swiftu~light curves, and they are events that tend to be intensely observed by a whole range of ground- and space-based facilities. Every few years, \swift~delights us with an even more remarkable example. These events tend to be bright, due to a combination of being intrinsically bright (isotropic gamma-ray energy, $E_{iso}$, $\sim$10$^{54}\,{\rm erg}$) and relatively nearby ($z\lesssim 1$), compared to the typical redshift of \swiftu-detected GRBs (see Figure \ref{fig:redshift_distribution}). 

The first markedly bright GRB was GRB 061007. At the time, it was the brightest observed by \swiftu,~and one of the brightest to be detected 
by \swift~ in BAT and XRT~\citep{sch07b,mun07}. This GRB was at a redshift of $1.26$ \citep{5716} and had an $E_{iso}\sim 1\times10^{54}\,{\rm erg}$ \citep{5722}. 
The afterglow had a $v$-band magnitude <11.1 at 80\,s, after the prompt emission \citep{sch07b}. This GRB was surpassed in brightness, a 
couple of years later, with the detection of GRB 080319B. This GRB was at a redshift of 0.937 \citep{7444} and had a V-band magnitude of 5.3 during the prompt emission. It was dubbed the `naked eye GRB' as it would have been visible to the naked eye for a few tens of seconds for an observer in a dark location looking at the right time and place. GRB 080319B saturated the \swiftu~$white$ filter for the first $\sim$970 s after the BAT trigger. From $\sim 280$s, a measurement could, however, be achieved in the $v$ filter.  For this GRB afterglow, a two-component afterglow model was necessary to explain the complex behaviour of the optical-X-ray emission \citep{racusin08}. The previous record holder for optical brightness was a pre-\swift~GRB, GRB 990123, with a ninth magnitude V-band~observation. 

GRB 1304027A was a $v$=12th mag, luminous ($E_{iso}\sim 8\times10^{53}\,{\rm erg}$; \citep[]{mas14}) and unusually close by ($z = 0.34$; \citep[]{14455}) GRB. Such a nearby and luminous event was predicted to occur once every 60 years \citep{mas14}. Long GRBs at $z<0.4$ tend to be sub-luminous ($E_{iso}<10^{52}\,{\rm erg}$, see also Section \ref{rare}), and so GRB 130427A was unusual in that its behaviour is more consistent with a GRB occurring at $z\sim$ 1--2. It was thus given the moniker 'a nearby ordinary monster' \citep{mas14}. GRB 160625B was another bright GRB. It triggered Fermi \citep{19580,19581}, but unfortunately not \swiftb,~and so the early optical behaviour was missed by \swiftu. Observations only began with \swiftu~9ks after the Fermi trigger \citep{tro17}. GRB 190114C had similar redshift and isotropic energy as GRB 130427A \citep{aje20}; it had a bright, 13th mag optical/UV afterglow, and was the first GRB to have TeV photons reported and detected in the range of \mbox{0.2--1~TeV}~\citep{mag19}. 

\begin{figure}[H]
\hspace{-3pt}\includegraphics[width=0.7\textwidth, trim={0cm 0.cm 1.5cm 0.5cm},clip]{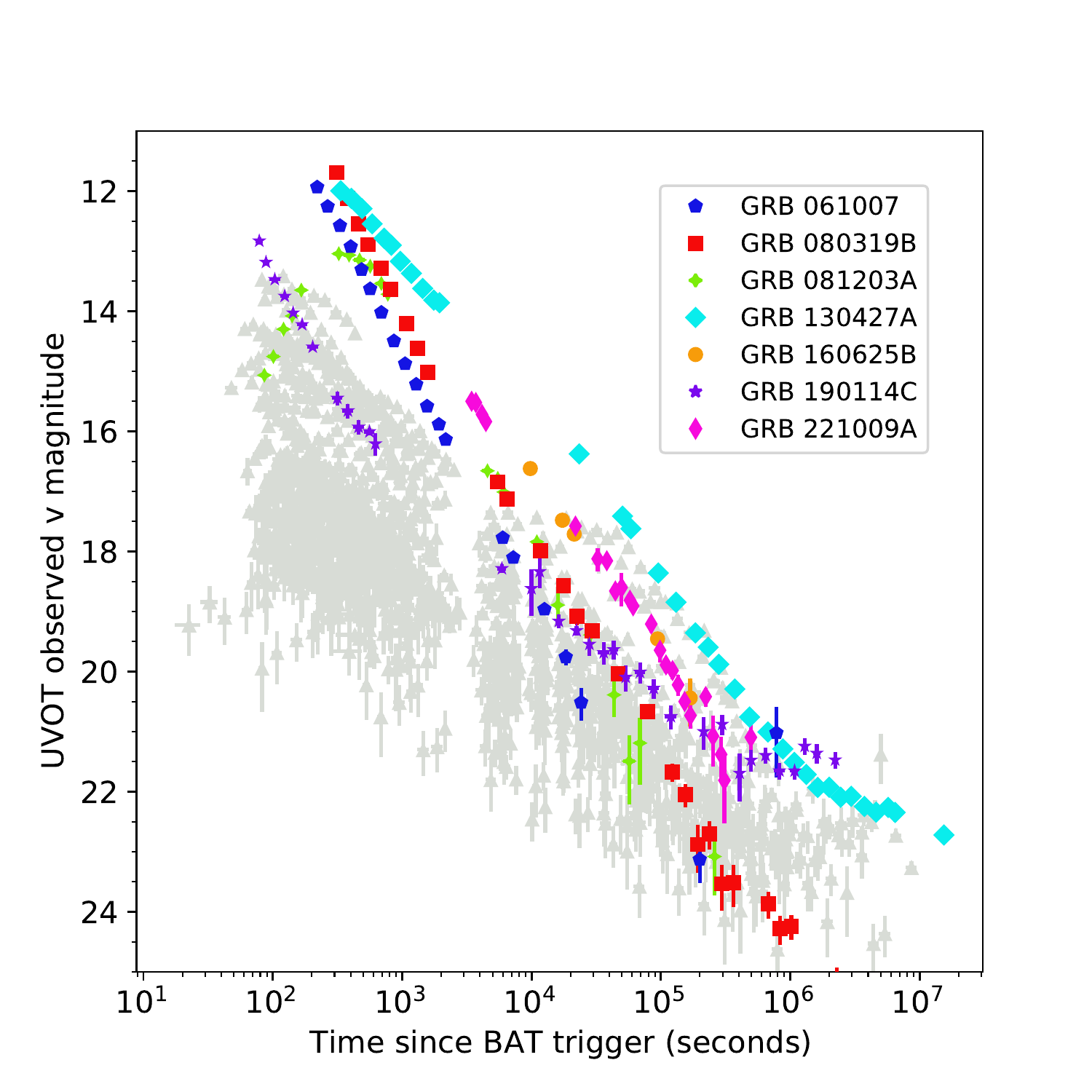}
\caption{A sample of long GRB UVOT light curves. Those in colour are the brightest observed by UVOT (see Section \ref{record}) and include GRB 061007, GRB 080319B, GRB 081203A, GRB 160625B, GRB 190114C, and GRB 221009A. The grey light curves are UVOT-observed GRBs discovered between 2005 and 2010, taken from \cite{oates12}. To create each light curve, individual colour filters ($b$, $u$, $uvw1$, $uvm2$, $uvw2$, and $white$) were normalised to the $v$-band light curve and were co-added to construct a single filter, $v$-band, and light curve. These light curves have not been corrected for extinction. For the light curves highlighted in colour, the extinction is small $A_v<0.5$ mag, except for GRB 221009A, which has a significant extinction with $A_v=4.1$ mag \citep{schl11}.}
\label{UVOTlcs}
\end{figure}

The latest addition to this collection of bright events observed by \swiftu~is the very recently detected GRB 221009A \citep{32632,32635}. This GRB was initially thought to be an X-ray transient due to its unusual gamma-ray behaviour observed by \swiftb,~and its apparent location within the Galactic plane. However, its status as a GRB was confirmed with the X-ray light curve decaying as expected for a GRB afterglow \citep{32635}, and subsequently, a redshift determination of 0.151 \citep{32648,32686,mal23}.
Most unusually, many facilities, including Fermi, AGILE, Konus-Wind, and INTEGRAL, reported detecting X-ray and gamma-ray emission from this GRB an hour prior to the \swift~detection \citep{32636,32637, 32650,32660, 32661,32668,32685,32751}. It is also the second GRB to be detected at TeV energies, and the first with photons of above 10 TeV \citep{32677}, with one photon being reported at 251~TeV \citep{ATel15669}. The isotropic energy from this event is still to be confirmed, but initial estimates suggest it is $E_{iso} \sim$ 1--6$\,\times\,10^{54}$ \citep{32660, 32668, 32762,kan23}, close to that of GRB 160625B~\citep{32762}. Since \swiftb~triggered on the afterglow emission (the first time in 18 years) rather than the prompt gamma-ray emission, \swiftu~observations only began an hour after the Fermi trigger \citep{wil23}. In addition, due to the GRB's closeness to the Galactic plane, the optical/UV emission is affected by significant galactic extinction ($A_v\simeq4$ mag increasing in the UV to an $A_{uvw2}\simeq10$ mag). Given these restrictions, GRB 221009A was still detected in the initial \swiftu~$v$ image at 16.6 mag, and marginally detected in $uvw1$. Such a nearby and bright event is only expected once in 1000 years \citep{wil23}. The results of the detailed analysis of this GRB are starting to be published; see \citep{ai22,alv22,car22, che22,gal22,gon22,lih22,mur22,nak22,ren22,rom22,rud22,rud22a,sat22, smi22,tro22a,var22,xia22,zha22,zhe22,zhu22,abb23, cam23, das23, fin23,ful23, gua23, kan23, las23,lev23,liu23, mal23,neg23,oco23,rip23,sah23, shr23, vas23, wil23,zha23,zha23b}. The \swiftu~light curves of all these bright GRBs are given in Figure \ref{UVOTlcs}.

\subsection{Rare and Unusual GRBs}
\label{rare}
As well as spectacular GRBs in terms of optical brightness, \swiftu~has observed GRBs with rare and unexpected behaviour.
Prior to \swift,~the nearest GRBs to us were GRB 980425 \citep{tin98,fol06} at $z=0.0085$, followed by GRB 031203A at 0.105 \cite{pro04,mar07}, and GRB 030329 at z = 0.168 \citep{2020,tho07}. GRB 980425 and GRB 031203 were of low luminosity, atypical of the bulk of cosmological GRBs. \swift~has increased the number of detected rare and nearby long GRBs, with GRB 060218 at z=0.033 \citep{4792}, GRB 100316D at z = 0.0591~\citep{sta11}, and GRB 111005A at z = 0.01326 \citep{mic18,tan18}. These GRBs were also of low luminosity, and along with GRB 980425 and GRB 031203, they are categorised as low-luminosity GRBs (LLGRBs). All of these events, except GRB 111005A \citep{mic18,tan18}, are associated with a type Ib/c supernova, supporting the connection between GRBs and the collapse of massive stars. Unfortunately, GRB 111005A was constrained by the Sun and thus was not observable with \swiftu,~and no variable optical or UV source was detected for GRB 100316D, which was hampered by the bright underlying host galaxy \citep{sta11}. However, \swiftu~did detect optical/UV emission from GRB 060218, with observations starting $\sim$100 s after the gamma-ray trigger. In this case, \swiftu~observations were important to constrain a thermal component, found initially in the X-ray spectrum, as it cooled and shifted with time into the optical/UV~\citep{cam06, pia06,eme19}. This was suggested to be the `shock break out' of the supernova, the moment where the shock wave breaks out of the star \citep{cam06, pia06}. This was the first time such an observation was observed with a GRB. 

Over the years, a small number of GRBs have been found by \swift~and other gamma-ray detectors to have \tninety~$>1000$ s \citep{gen13,lev14,kan18}, including GRB 060218 and GRB 100316D. An even smaller number are ultra-long GRBs (ULGRBs), with \tninety~$>2000$ s \citep{gen13,lev14}. Of these, \swift~detected: GRB 101225A at $z = 0.847$ \citep{tho11,cam11,lev14}, GRB 111209A at \mbox{$z = 0.677$}~\mbox{\citep{12648,gen13,str13,kan18}}, GRB 121027A at $z = 1.773$ \citep{lev14}, and GRB 130925A at $z = 0.347$ \citep{15249,eva14,gre14} (other candidate ULGRBs are discussed in \cite{kan18}). 
While all four were observed by \swiftu, only GRB 101225A and GRB 111209A have well-sampled \swiftu~light curves. GRB 130925A was not detected by \swiftu,~and GRB 121027A was only weakly detected in $white$, which may be due to its larger distance compared to GRB 101225A and GRB 111209A~\citep{13932,lev14}. The main difficulty in determining the nature of ULGRBs is in explaining their long gamma-ray duration. It has been suggested that they are caused by the collapse of massive stars with radii that are larger than that considered for typical GRB progenitors \citep{gen13,lev14}. An alternative suggestion is that these are the tidal disruption of white dwarfs as they pass close to a supermassive black hole (SMBH), for which the black hole is at the lower mass end of the SMBH distribution (e.g., <$10^5M_\odot$)~\citep{lev14}. An analysis of the \swiftu~and GROND data of GRB 111209A by \cite{kan18} suggests that the optical/UV afterglow is consistent with the main population of long GRBs in terms of its luminosity distribution. They also state that while GRB 111209A has an isotropic energy and peak energy at the high end of the GRB distribution in terms of prompt emission parameters, these parameters are consistent with the Amati and Ghirlanda relations \citep{ama02,ghi04}, suggesting that this and potentially other ULGRBs may not be a distinct class. Ref. \cite{kan18} also notes that GRB 101225A is more unusual in its optical/UV behaviour compared to the other ULGRBs, and may have a distinctive progenitor. GRB 101225A has been explained by \cite{tho11} as an inspiral of a neutron star into a helium star, creating a central engine similar to a GRB.

\subsection{Dark GRBs}
Before moving on to discuss what we have learned about the optical/UV behaviour of GRBs, it is important to discuss the large fraction of long GRBs that were observed by \swiftu~but not detected, $\sim$40--60\%, depending on the detection threshold (see \mbox{Section \ref{uvot_overview}}). The optical/UV afterglows of some GRBs may not be detected due to observational constraints such as the rapidity of telescope pointing; and the GRB location relative to the Sun, Moon, and other bright sources. Restricting the sample to those observed by \swiftu~within the first 500 s after the gamma-ray detection and without observing constraints, there is still a significant fraction that are `dark'. Formally, GRBs are considered to be ``dark'' if their X-ray to optical spectral index is $\beta_{OX} <\beta_X-0.5$ \citep{van09}, and so, some GRBs may be considered to be `dark', even with the detection of the optical/IR emission. The percentage of `dark' GRBs typically found for different instruments is 20--50\% \citep{mel08, cenko09, zhe09, gre11}. With the rapid slewing capabilities of \swift, observations with \swiftu~have enabled us to rule out these dark GRBs as being due to factors such as a lack of sensitivity, late observation times, and rapid temporal decays \citep{mel12}. The low detection rate or ``darkness'', could however, be due to one or more of the following: a high background due to a small Sun-to-field angle~\citep{fyn09}, a large galactic extinction (e.g., \citep[]{fyn01,fyn09}), a high circumburst extinction (e.g., \citep[]{rom06b,del11,jeo14}), intrinsic faintness \citep{del11}, and Ly$\alpha$ damping due to high redshift (e.g., \citep[]{rom06b,del11,chi19}). Several studies, including those using \swiftu~observations (e.g., \citep[]{kru11}), have suggested that high circumburst extinction and high-redshift are the two main causes for why GRBs are `dark'~\citep{per09, zhe09, gre11, mel12}.

\subsection{The Optical/UV Behaviour of \swiftu~Observed GRBs}
\swiftu~has provided the GRB community with the largest sample of optical/UV observations of GRBs to date, with observations beginning typically 120 s after the detection of the gamma-ray emission. We now have a good idea of how GRBs behave in the optical/UV range from the first few minutes after detection, through to the first few days, beyond which they typically fade below the detection sensitivity of \swiftu. 

Since X-ray and optical/UV observations are taken simultaneously, it is useful to compare the temporal behaviour in both bands. Not long after the launch of \swift, the X-ray light curves were shown to have two unexpected features: an initial steep decay, thought to be due to the tail of the prompt emission, and a shallow decline phase whose nature is not yet resolved, but may be due to a long lived central engine \citep{zhang01} or a short lived central engine that emitted shells over a wide range of velocities, with the slower shells catching up at a later time to continuously eject energy into the ejecta \citep{ree98}. These two power-law segments, in addition to the two power-law segments observed pre-\swift, the normal decay and the post-jet break decay, lead to the creation of a `canonical' X-ray light curve, whereby most X-ray light curves can be fitted with one or more of these power-law decays, and X-ray flares may also be additionally superimposed \citep{nousek,zhang06}. 

\begin{figure}[H]
\vspace{-3pt}
\includegraphics[width=0.9\textwidth]{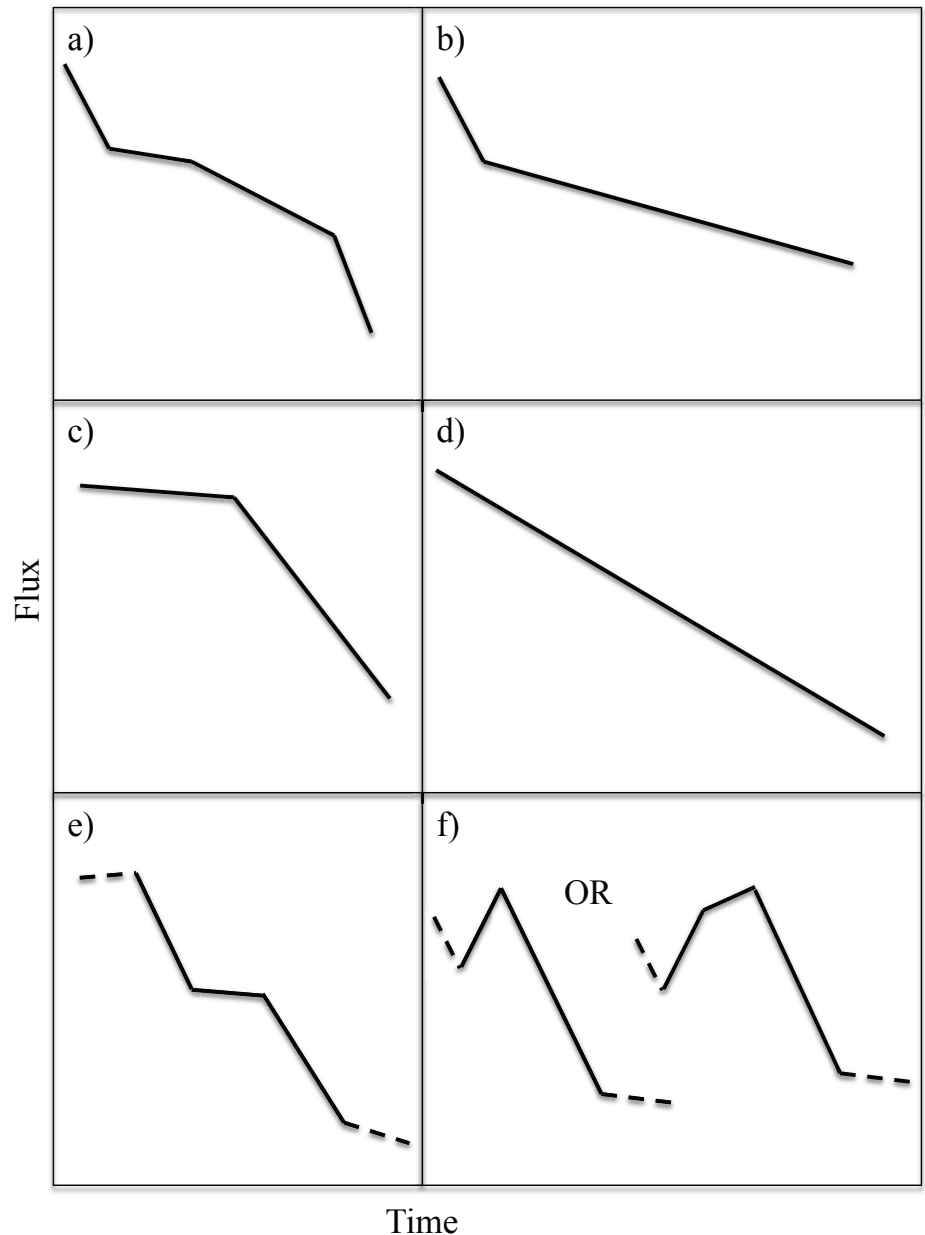}
\caption{Schematics of GRB optical/UV light-curve morphologies, as derived from \cite{rom17}. Morphologies (\textbf{a}--\textbf{d}) are the same as those used to describe the X-ray light curves in \cite{eva09}, but morphologies (\textbf{e}--\textbf{f}) are specific to the optical/UV. These additional two morphologies represent $25\%$ of the \swiftu~GRB light curves. Dotted lines represent those portions of the light curves that are not always seen in these morphologies. The percentage of \swiftu~light curves corresponding to the different panels are: (\textbf{a}) 7\%, (\textbf{b}) 7\%, (\textbf{c}) 14\%, (\textbf{d}) 47\%, (\textbf{e}) 21\%, and (\textbf{f}) 4\%. This figure is reproduced from Figure 4 of \citep{rom17}. 
 \copyright AAS. Reproduced with permission.}
\label{fig:UVOTcatlc}
\end{figure}

In the optical/UV, it was shown that the optical/UV light curves behave more simply~\citep{oates09}, with an initial study of optical/UV light curves observed by \swiftu,~showing that the light curves decay as a simple or broken power-law, and with a small number displaying an initial rise to a peak before they decay. Later, with a larger sample of GRBs, the optical/UV light curves could be grouped into different morphologies \citep{rom17} in a similar fashion, as was achieved for the X-ray light curves \citep{eva09}. The morphologies are shown in the eight panels of Figure \ref{fig:UVOTcatlc}, (a) a `canonical’ light curve, (b) a break to a shallower decline, (c) a break to a steeper decline, (d) no break, (e) a gentle rise, transitioning to a steep decay, followed by a shallow decay, then another steep decay, and finally a more gentle decay; (f)~starts with a rapid, steep decay, then a rise to peak (in some instances, with a break within the rise), a steep decay, and a final less-steep decay. Panels (a--d) are the same as those used for the X-ray light curves, while (e) and (f) are specific to the optical/UV emission. The final shallow decay in (f) is likely a result of poor background subtraction due to the background host signal dominating over the GRB signal. Using this scheme, approximately half of the optical/UV afterglows observed by \swiftu~are consistent with panel (d), a simple power-law, 21\% are consistent in behaviour with the scheme in panel (e), with a few percent being consistent with each of the other panels. In comparison for the X-ray light curves, 42\% are consistent with panel (a) canonical, 30\% are consistent with one break, 15\% with panel (b) and 15\% with panel (c), 4\% are consistent with panel (d) a simple power-law, and a further 24\% are considered to be oddballs, displaying a range of behaviour that is not consistent with these specific types. This morphological categorisation provides some indication of the number and frequency of emission components producing the X-ray and optical light curves. For the optical/UV, at least $\sim$50\% are consistent with panel d), which suggests that typically, a single emission component is sufficient to produce all of the optical/UV emission. For the X-ray light curves, the largest fraction of GRBs are consistent with panel (a), suggesting a more complex scenario with multiple emission components producing the X-ray emission, as described in the previous paragraph. However, ref. \cite{rom17} states that poorly sampled light curves tend to be consistent with a power-law, and so light curves with poor sampling may have more complex behaviour than that observed.

The range in optical/UV behaviour can also be displayed as a canonical optical light curve \citep{li12} in a similar way as was done for the X-ray canonical light curve \citep{nousek,zhang06}, with the optical/UV afterglow comprised of or more components: prompt optical flares, an early optical flare from the reverse shock, shallow-decay segment, the standard afterglow component (an onset hump followed by a normal decay segment), the post-jet-break phase, optical flares, rebrightening humps, and late supernova bumps \citep{li12}. I will now focus on the particular features observed in the optical/UV light curves.

\subsubsection{Early Optical Emission}
Some of the most informative and important observations \swiftu~has performed have been in relation to the earliest phases of the optical/UV GRB emission, within 
the first few hundred seconds. Optical/UV emission, when detected, is observed from the earliest moment that \swiftu~observations begin \citep{pag19}. In the majority 
of cases, the optical/UV emission is already declining by the time \swiftu~observations begin, and only around 20\% are observed to have an initial rise to a peak in 
their light curves \citep{oates09,rom17,pag19}. A correlation has been observed by \cite{lia10,pan11,lia13} between the peak time and the peak brightness of optical afterglows, 
but for the majority of GRBs, we do not have a measure for either parameter. Since all of the optical/UV light curves must rise initially, this suggests the need for an even 
faster response to catch the start of the optical/UV emission. 

There are two possible origins for the early optical/UV emission. It could either be produced by the external shocks that are generated as the ejecta are slowed 
by the external medium, or the optical emission could be related to prompt gamma-ray emission, which in turn is thought to be produced internally to the outflow. 
The observed emission may also be a combination of both possibilities, with different components dominating at different times for different GRBs. For the majority of GRBs, 
the prompt emission is not thought to be the dominant emission mechanism producing the early optical/UV emission of \swiftu~observed GRBs, but it is 
observed in some \citep{kop13,pag19}. Examples where the prompt emission contributes to the early optical/UV emission observed by \swiftu~are GRB 061121 \citep{pag07,oga19}, 
GRB 080928 \citep{ros11,oga19}, GRB 110205A \citep{oga19}, and the ULGRB GRB 111209A \citep{str13, kan18}. In addition, prompt emission produced some of the very early optical emission of GRB 080319B \citep{racusin08}, before \swiftu~began observations. Simultaneous observations of the prompt emission across multiple wavelengths are 
rare but important for understanding the origin of the prompt emission and the transition to the afterglow phase \citep{oga19}.

\begin{figure}[H]
\vspace{-3pt}
\begin{tabular}{cc} 
\begin{subfigure}[b]{0.46\textwidth}
\hspace{-6pt}\includegraphics[width=0.89\textwidth]{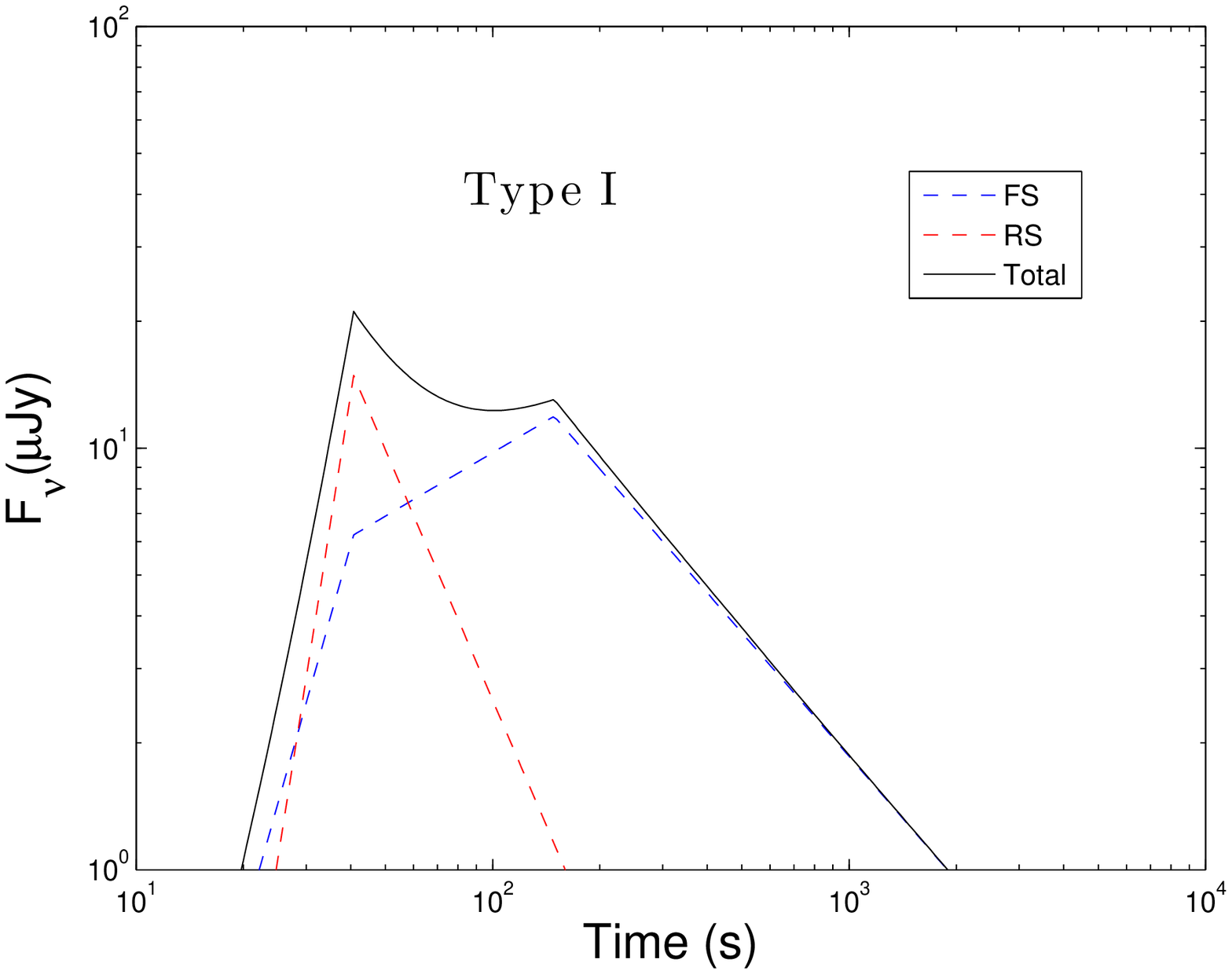}
\end{subfigure}&
\begin{subfigure}[b]{0.46\textwidth}
\includegraphics[width=0.89\textwidth]{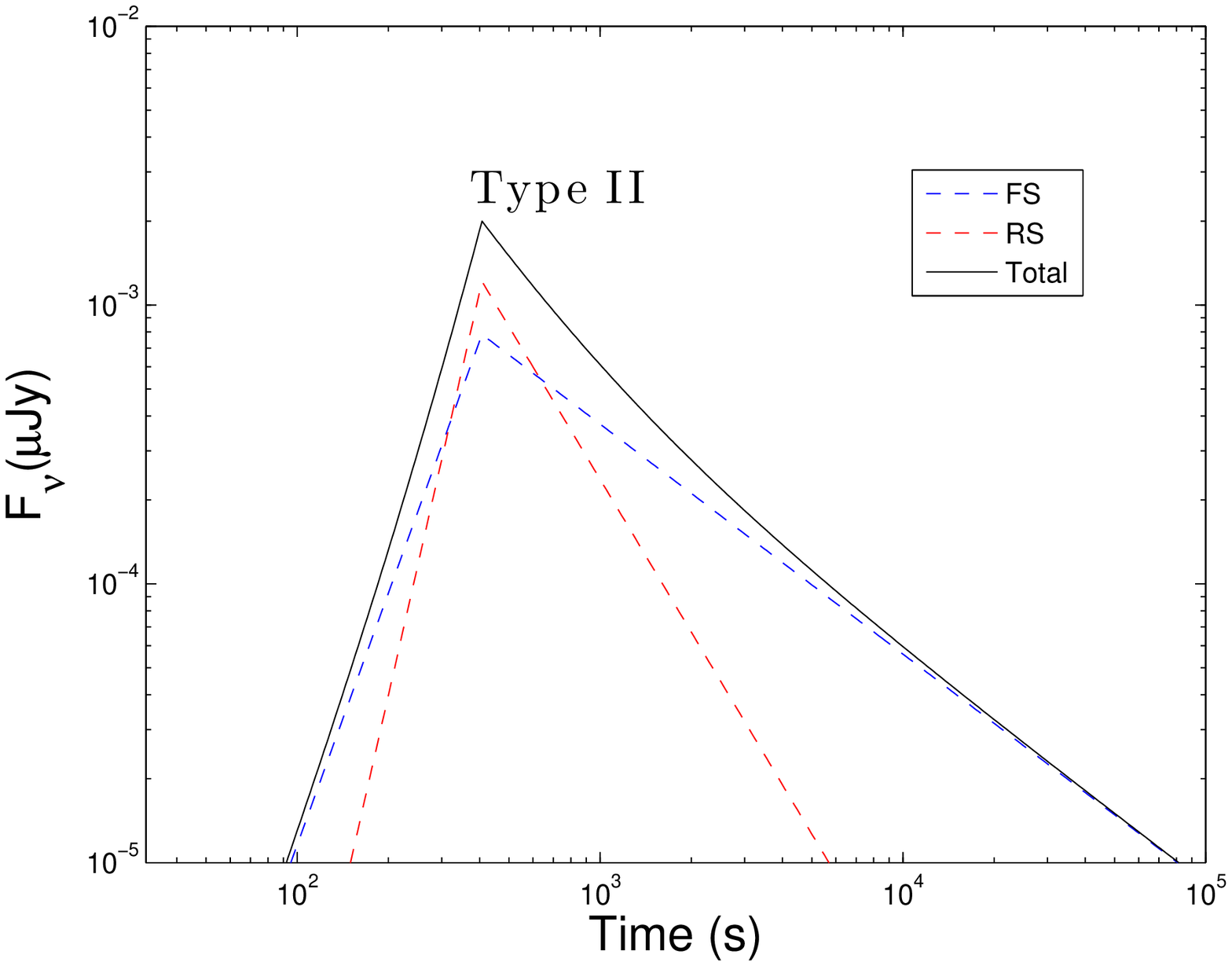}
\end{subfigure}\\
\begin{subfigure}[b]{0.46\textwidth}
\hspace{-6pt}\includegraphics[width=0.89\textwidth]{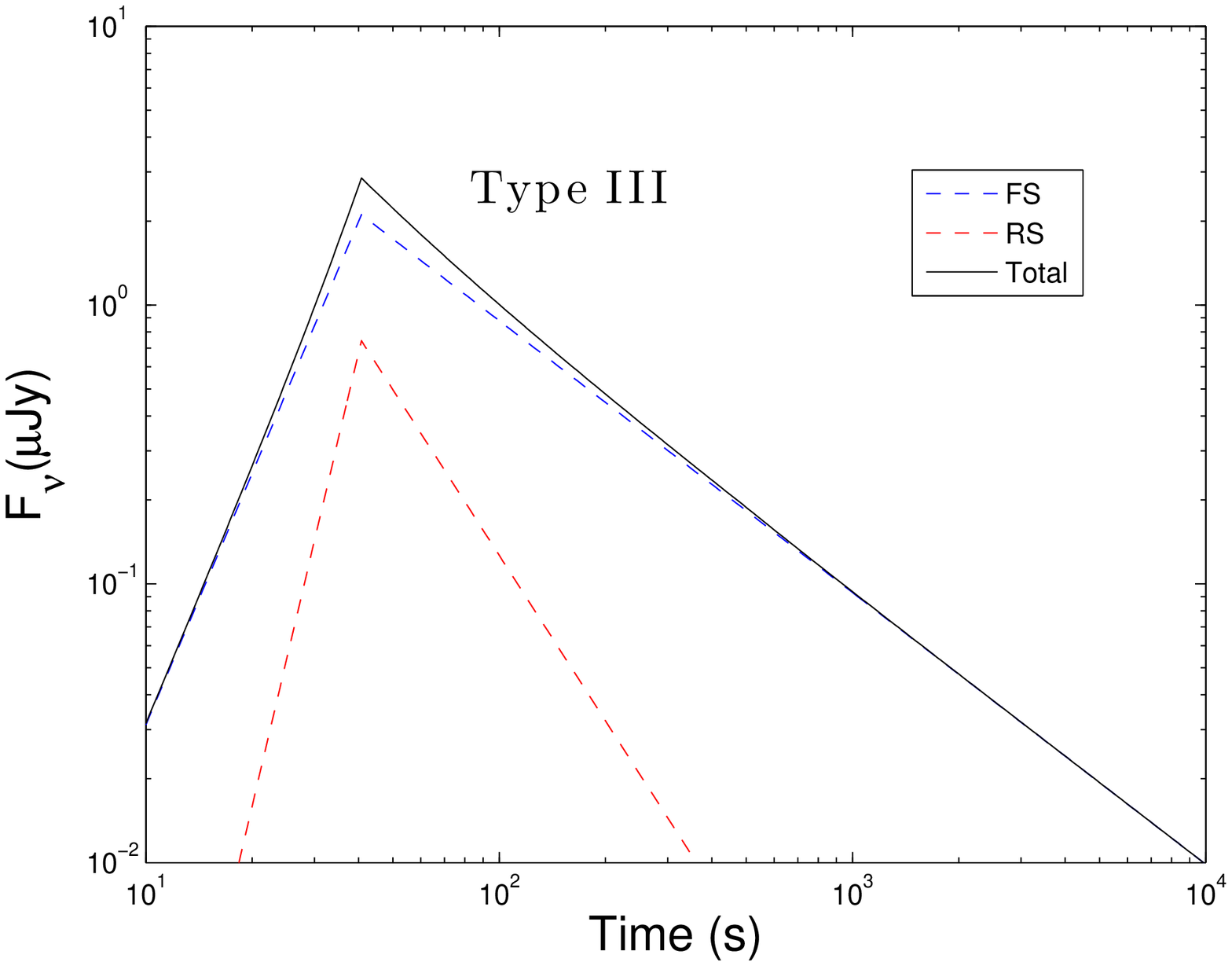}
\end{subfigure}&
\begin{subfigure}[b]{0.46\textwidth}
\includegraphics[width=0.89\textwidth]{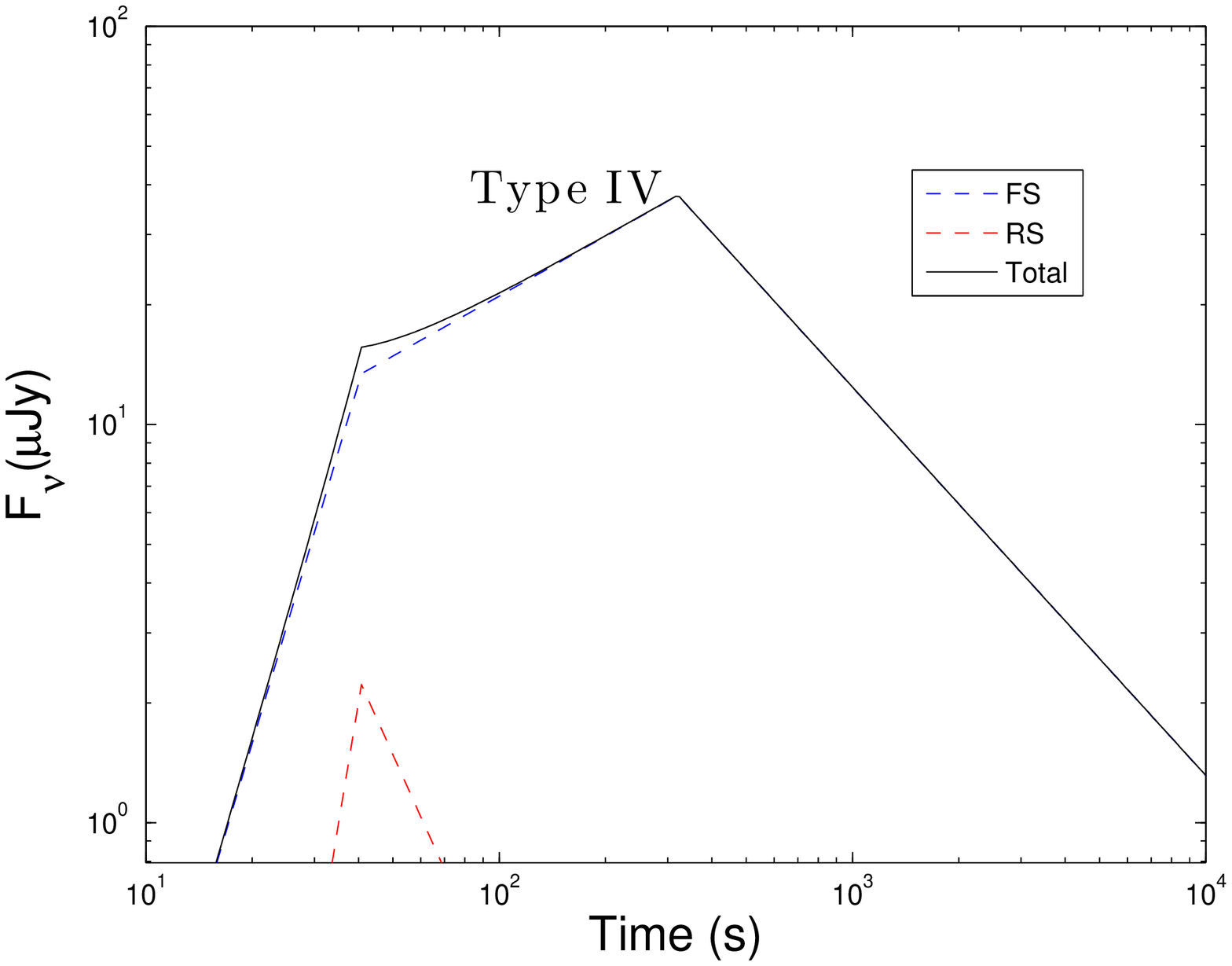}
\end{subfigure}
\end{tabular}
\caption{Example optical light curves for four different scenarios of early afterglow behaviour resulting from the combination of forward and reverse shock emission \citep{gao15}. Top left panel (\textbf{type~1}): the light curve is initially dominated by the reverse shock (RS) emission, but the reverse shock quickly fades and a second peak is produced as $\nu_m$ crosses the optical band in the forward shock (FS). Top right (\textbf{type II}): the forward shock and reverse shock peak at the same time, but the reverse shock dominates the emission at the peak. The reverse shock quickly fades and the decaying forward shock emission dominates at late times. Bottom left (\textbf{type III}): the forward shock always dominates; the reverse shock is weak or suppressed. The peak is due to the onset of the forward shock emission. Bottom right (\textbf{type IV}): the forward shock dominates, and the reverse shock is weak or suppressed. The onset of the forward shock emission produces the first (lower) peak; the main peak occurs when $\nu_m$ crosses the optical band. Figures reproduced from Figure 1 of \citep{gao15}. \copyright AAS. Reproduced with~permission.}
\label{early_afterglow}
\end{figure}

For the small fraction of GRBs with early observed rises in their optical/UV light curves, there are a number of mechanisms that may produce it. In some cases, an initial 
rise may be due to the prompt emission \citep{kop13}. Observed rises may also be produced by the external shock. Both the forward and reverse shocks could produce 
a rise in the optical/UV afterglow, and may even be a combination of both \citep{zhang03,zhang05,gom09,gao15}; see Figure \ref{early_afterglow}. In the forward shock, a rise 
may be produced as a result of the jet slowing down as it ploughs into the external medium (the onset of the afterglow) or the passage of $\nu_m$ through the optical bandpass \citep{sari99, zhang03}. If the rise is due to the start of the forward shock, it should also be observed in the X-rays, but it is generally masked by the tail of the 
prompt emission~\citep{oates11,wil17}. In a sample of \swiftu~optical/UV light curves with observed initial rises, the passage of $\nu_m$ could be excluded as the 
cause of the initial peak, and there was no evidence for reverse shock emission \citep{oates09}. The rise in these optical/UV light curves and other early optical 
light curves samples suggested that the rise could be attributed to the start of the forward shock \citep{mol06,oates09,ryk09}. More recent work \citep{gao15}, with a larger 
sample of GRBs suggests that optical rises are produced by all possible combinations of forward and reverse shocks, as shown in Figure \ref{early_afterglow}, with 
$\sim 50\%$ of rises being consistent with the start of the forward shock (bottom left of Figure \ref{early_afterglow}). The least common type of rise was initially 
dominated by the reverse shock that quickly fades, and then a second peak occurs due to the passage of $\nu_m$ through the optical band of the forward shock \citep{gao15}. It is possible, however, that some of these optical rises occur while the prompt emission is still active, and so they could also be explained as internal origin \citep{kop13}.

For optical rises produced by the onset of the afterglow, the peak time of the optical light curve depends on whether the shell that collides
with the surrounding interstellar medium is thick or thin. For thick shells, the peak time, $t_{peak}$, of the optical emission will be comparable to \tninety~\citep{sar99,kob00}. In the thin shell case, the optical peak is expected after \tninety~\citep{mol06,ryk09}. Providing that the peak occurs when $t_{peak}>$\tninety~(thin shell regime), then the Lorentz factor of the shell at the moment of the peak, $\Gamma_{peak}$, can be derived \citep{mol06,sari97,sari99}; $\Gamma_{peak}$ 
is expected to be half of the initial value $\Gamma_0$ \citep{pan00,mes06}. A lower limit can be obtained for those GRBs where the peak occurs before the onset of observations ($>$100; \citep[]{oates09}). For GRBs with observed rises, $\Gamma_0$ ranges from 100 to 1000 \citep{mol06,oates09,ryk09,lia10}, consistent with expectations that $\Gamma_0$ must be at least 100 in order to produce gamma-ray emission \citep{fen93,pir04}. Using $\Gamma_{peak}$, it is possible to deduce two more quantities: the isotropic-equivalent mass of the baryonic load, $M_{fb}$, and the deceleration radius, $R_{dec}$ \citep{mol06}. The deceleration radius defines the radius, in a thin shell regime, at which the accumulated external medium is $1/\Gamma_0$ of the ejecta mass. The mean mass of the baryonic load and the mean deceleration radius for GRBs with an observed rise is $\langle{M_{fb}}\rangle\sim 4\times 10^{-3}\,{\rm M}_{\odot}$ and $\langle{R_{dec}}\rangle\sim 2\times 10^{17}$~cm. For GRBs without observed rises, the quantities are $\langle{M_{fb}}\rangle \lesssim 1\times 10^{-3}\,{\rm M}_{\odot}$ and $\langle{R_{dec}}\rangle\lesssim1\times 10^{17}$~cm~\citep{oates09,lia10}. These deceleration radii are consistent with the value expected from the forward shock model, $R_{dec}\sim10^{16}$~cm \citep{ree92}. The internal shocks, that are believed to power the prompt emission, are expected to occur at $\sim$$10^{15}\,{\rm cm}$ \citep{mes97,rees94}.

\subsubsection{Optical Flares} 
\begin{figure}[H]
\includegraphics[width=0.9\textwidth]{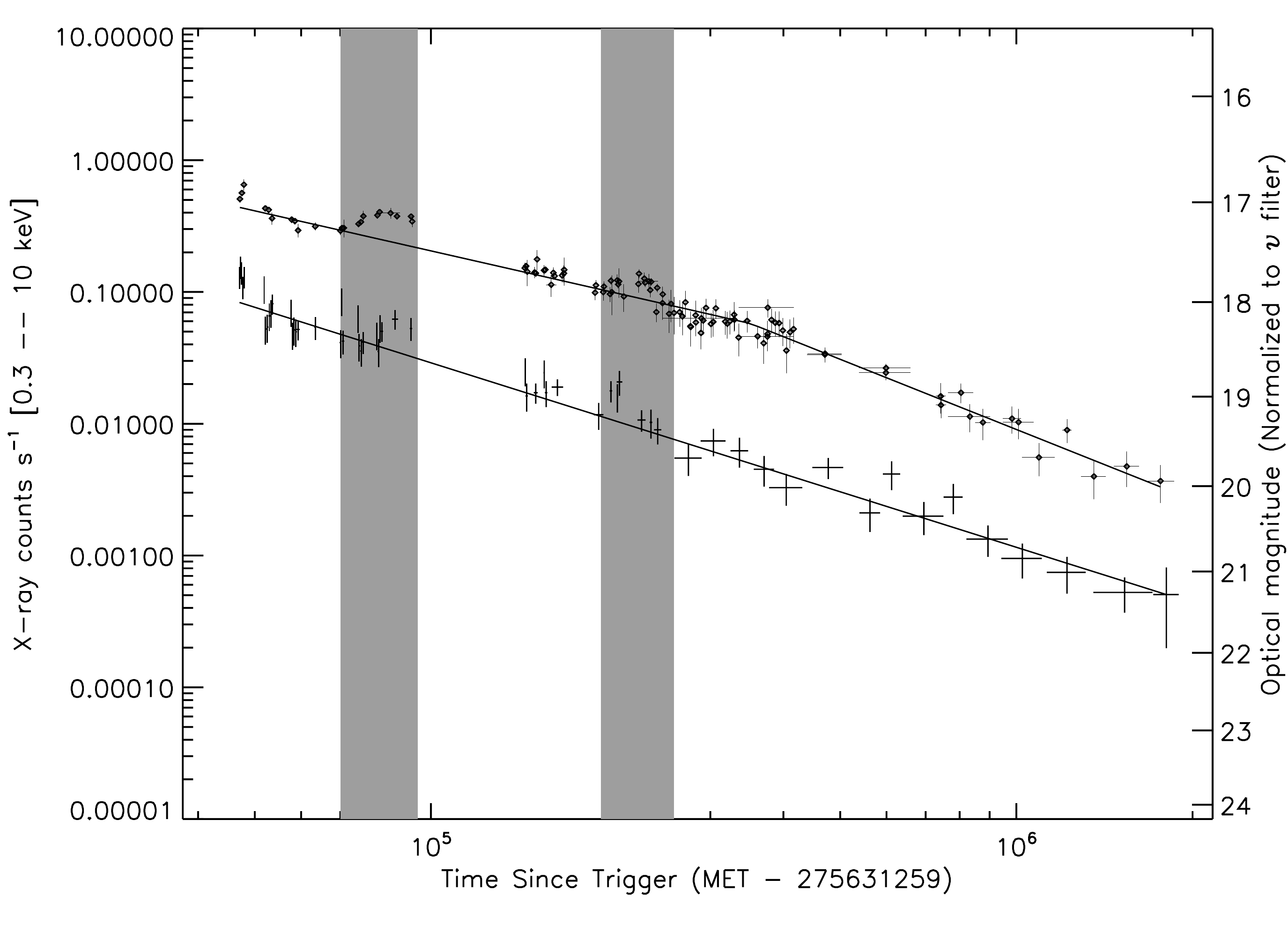}
\caption{X-ray and optical/UV light curves of GRB 090926A. Flaring periods are shaded in grey. The \swiftu~light curve is given at the top (circles); the XRT light curve is below (pluses). Figure reproduced from Figure 2 of \cite{swe10}. 
\copyright AAS. Reproduced with permission.}
\label{uvot_flare}
\end{figure}

 Flares in X-ray light curves had been seen prior to \swift, but only a handful of times (e.g., \citep[]{pir98,pir05,gal06a}). With the launch of \swift, it was quickly shown that they are quite common \citep{bur05b,rom06}, appearing in approximately 50\% of XRT afterglows \citep{obr06}, occurring generally after the end of \tninety,~and were superimposed on the X-ray light curve. These X-ray flares are thought to be due to the same processes that produce the prompt gamma-ray emission \citep{fal07,chi07}. Flares are also observed in the optical/UV light curve (see example in Figure \ref{uvot_flare}). However, they are generally not as prominent as those in the X-rays, not as frequent \citep{li12}, and were likely to be overlooked or dismissed as noise \citep{swe13}. 
 
 Using a blind, systematic search for flares, ref. \cite{swe13} analysed 201 \swiftu~GRB light curves, and found episodes of flaring in 68 ($\sim$34\%); a lower fraction than that found in X-ray light curve samples ($50\%$), but higher than previous optical studies (12\% \citep[]{li12}). On average, two flares were found in each of the 68 GRBs with flares. Ref. \cite{swe13} found that most of the flares occur within the first 1000 s of the afterglow, but could be observed and detected beyond $10^5$ s. More than 80\% of the flares detected are short in duration, with $\Delta t/t$ of $<$0.5. Ref.~\cite{yi17} investigated further the \swiftu~optical flares found by \cite{swe13}. Ref. \cite{yi17} found a correlation between the rise and decay times of the optical flares, and a correlation between their duration and peak time. These correlations are consistent with the results of X-ray flares, suggesting they share the same physical origin, and both being possibly produced via internal emission as a result of central engine activity \citep{yi17}.

\subsubsection{Optical Rebrightenings}
Another feature found in the \swiftu~observations are late optical/UV rebrightening bumps (e.g., GRB 100815A; \citep[]{nar14,depas15}). These have been detected in around 18\% of optical light curves \citep{lia13}. These bumps occur after the initial onset and detection of the optical emission, and they are distinctive from bumps produced by an accompanying supernova. Supernovae occur later, peaking around one to two weeks after a GRB trigger, and they are generally red in colour. While some rebrightening bumps can be explained through a combination of reverse and forward shocks (e.g., the top left panel of Figure \ref{early_afterglow}; \citep[]{depas15}), many require a structured jet \cite{lia13}. In this case, the jet is not a simple uniform jet, but the energy and Lorentz factor of different parts of the jet varies and is dependent on the angle away from the jet axis. The simplest structured jet is a two-component outflow, which has an inner, narrow jet and an outer, wider jet encompassing the narrow jet. The narrow jet produces the prompt emission together with the initial optical emission, and the wider outflow dominates the optical emission at later times \citep{granot02}. This model interprets the rebrightening bump as the deceleration of the second slow jet. A two-component outflow model has a great deal of flexibility. Variations on this model have been used to explain the behaviours of GRBs with breaks in the X-ray light curves, which are geometric in nature (e.g., no spectral evolution is observed at the time of the X-ray break; spectral evolution would have indicated that the break was due to the passage of a synchrotron frequency through the observing band), but they have no corresponding break in the optical/UV light curves, which would otherwise be expected in the uniform jet model (e.g., \citep[]{oates07,depas09, racusin08}).

\subsubsection{Jet Breaks}
\textls[-25]{Jet breaks were thought to be observed in pre-\swift~observations of GRBs (e.g.,~{\citep[]{fra01,blo01,zeh06}})}, however, only a few were confirmed as being achromatic \citep{kul99,har01,klo04}. While there have been some cases of confirmed achromatic jet breaks in post-\swift~GRBs (e.g., GRB 050525A and GRB 140629A; \citep[]{blu06,hu19}), they are small in number \citep{bur06,liang08,eva09,racusin09}, and for some GRBs, no jet break has been detected for months \citep{gru07} or even years after the initial gamma-ray detection~\citep{depas15}. These extreme cases can pose problems for the external shock models, requiring extreme values of the physical parameters of the explosion, the emission mechanism, and the environment to explain them \citep{depas15}. In general, however, the lack of detected jet breaks for many GRBs may likely be, at least in part, due to the lack of good coverage, particularly in the optical/UV. Of 900 GRBs studied by \cite{wan15}, only 85 had well-sampled optical and X-ray afterglows. Within these 85, ref. \cite{wan18} found that around half had achromatic breaks, consistent with being a jet break \citep{wan18}, with a wide range of jet break time, from a few hundred seconds, up to $250\,{\rm ks}$. When a jet break can be determined, it can be used to constrain the opening angle of the jet, with values found to range from $\sim$$1^\circ$ to $\sim$$50^\circ$ (e.g.,~\citep{tan18,zha20}, see also references therein) and a typical value $\theta_j=(2.5\pm1.0)^\circ$~\citep{wan18}. The jet opening angle can be used to compute the geometrically corrected energy of a GRB as the energy of a GRB is concentrated with a jet and not emitted isotropically. The authors of \cite{wan18} compute a beaming-corrected gamma-ray energy $\log\,E_\gamma=(49.54\pm 1.29)\,{\rm erg}$ and a geometrically corrected kinetic energy ($E_K$), the blastwave kinetic energy is computed from the optical afterglow, $\log\,E_{K}=(51.33 \pm0.58)\,{\rm erg}$. With $E_\gamma$ and $E_K$, the radiative efficiency of the jet can be computed \citep{zhang07, racusin11}. Most GRBs in \cite{wan18} have a small radiative efficiency of $<$10\% at the time of the jet break. Determining accurate values for the $\theta_j$ and $E_\gamma$, as well as the time of the jet break, $t_{jet}$, are important for GRB correlations (e.g., \citep[]{fra01,ghi04,lia05}). The opening angle of the GRB jets, $\theta_j$, is also important in constraining the total GRB event rate density, with~ref. \citep{wan18} finding twice that of pre-\swift~estimates \citep{fra01,gue05}.

\subsubsection{Afterglow Luminosity and Correlations}
Soon after the launch of \swift, several works found a bimodal optical luminosity distribution \citep{nard06,nard08,liang}, implying two populations of optically bright GRBs. However, other studies \citep{mel08,cenko09,kan10,kan11, zan13}, including one study consisting only of \swiftu~light curves~\citep{oates09}, only required a single population to describe the optical luminosity distribution. A single population seems to best represent the distribution even as the sample continues to increase in size (e.g., \citep[]{kan20}).

Studies of single GRBs provide exceptional detail on the behaviour and physical properties of individual events. However, statistical investigations of large samples of GRBs aim to find common characteristics and correlations that link individual events and that therefore provide insight into the mechanisms common to GRBs. Statistical investigations have benefited greatly post-\swift~launch from the observation of large numbers of GRBs with well-sampled X-ray and optical/UV observations. This has led to a number of correlations being discovered in the \swift~era within the afterglow emission and linking the prompt gamma-ray emission \citep{lia05,kan10,bern12,mar13,zan16}. This is most notable between the intrinsic afterglow brightness and the isotropic energy of the prompt emission \citep{kou04,depas06,nys09,dav12,mar13,oates15}, which suggest the most energetic GRBs in the prompt emission have also the brightest afterglows. Since the isotropic energy of the prompt emission is correlated with the intrinsic prompt emission peak energy \citep{ama02,ama06}, this also implies a three-parameter correlation between the isotropic energy of the prompt emission, the intrinsic prompt emission peak energy, and the X-ray afterglow brightness \citep{bern12,mar13,zan16}. The intrinsic brightness of the afterglow in the X-ray and optical light curves are also correlated, such that GRBs with bright X-ray afterglows have also bright optical afterglows \citep{oates15}, see also \citep{jak04,geh08,ber14}. The Liang--Zhang correlation is another three-parameter correlation between the isotropic energy of the prompt emission, the intrinsic prompt emission peak energy, and the jet break time \citep{lia05}. For recent reviews describing all GRB correlations in the prompt and afterglow emission, see \cite{wan15b,dai17,dai18} and references therein. For a recent, comprehensive, and systematic study of all GRB parameters and correlations, see \citep{wan20}. Three correlations have been discovered using only optical/UV afterglow observations, which I will expand upon below. 

Refs. \cite{pan08,lia10,pan11,pan13} find a significant correlation between the peak time and peak afterglow brightness in both the X-ray and optical light curves of those GRBs with observed rises. Ref. \cite{lia10} used the peak time to determine the initial Lorentz factor of the outflow, finding it to be correlated with the isotropic gamma-ray energy.

Within samples of optical light curves \cite{li12, dai20, dai22} a correlation is found between the restframe time at the end of the plateau phase, with the luminosity at the same time (see also \citep{pan11}). This correlation is also present in the X-ray light curves \citep{dai08,ghi09,dai10,dai13}, and in GeV light curves \citep{dai21}. A theoretical interpretation of this correlation is that it may be explained within the context of the standard fireball model through the evolution of the microphysical parameters \citep{van14,van14b}. Alternatively, it may be explained by the spindown of a newly born magnetar \citep{dai98, zhang01,met11,bern12,wan22}, or by structured jets viewed over a range of viewing angles~\citep{ben20}. An extension of this optical luminosity-plateau end time correlation (also known as the Dainotti relation) has been obtained by adding the peak prompt luminosity, leading to a three-parameter 'Fundamental Plane' correlation \citep{dai22}, which was also shown to be present in the X-ray light curve sample \citep{dai16,dai17b}.

\begin{figure}[H]
\includegraphics[width=0.45\textwidth,trim={0cm 0cm 2.5cm 2.0cm},clip]{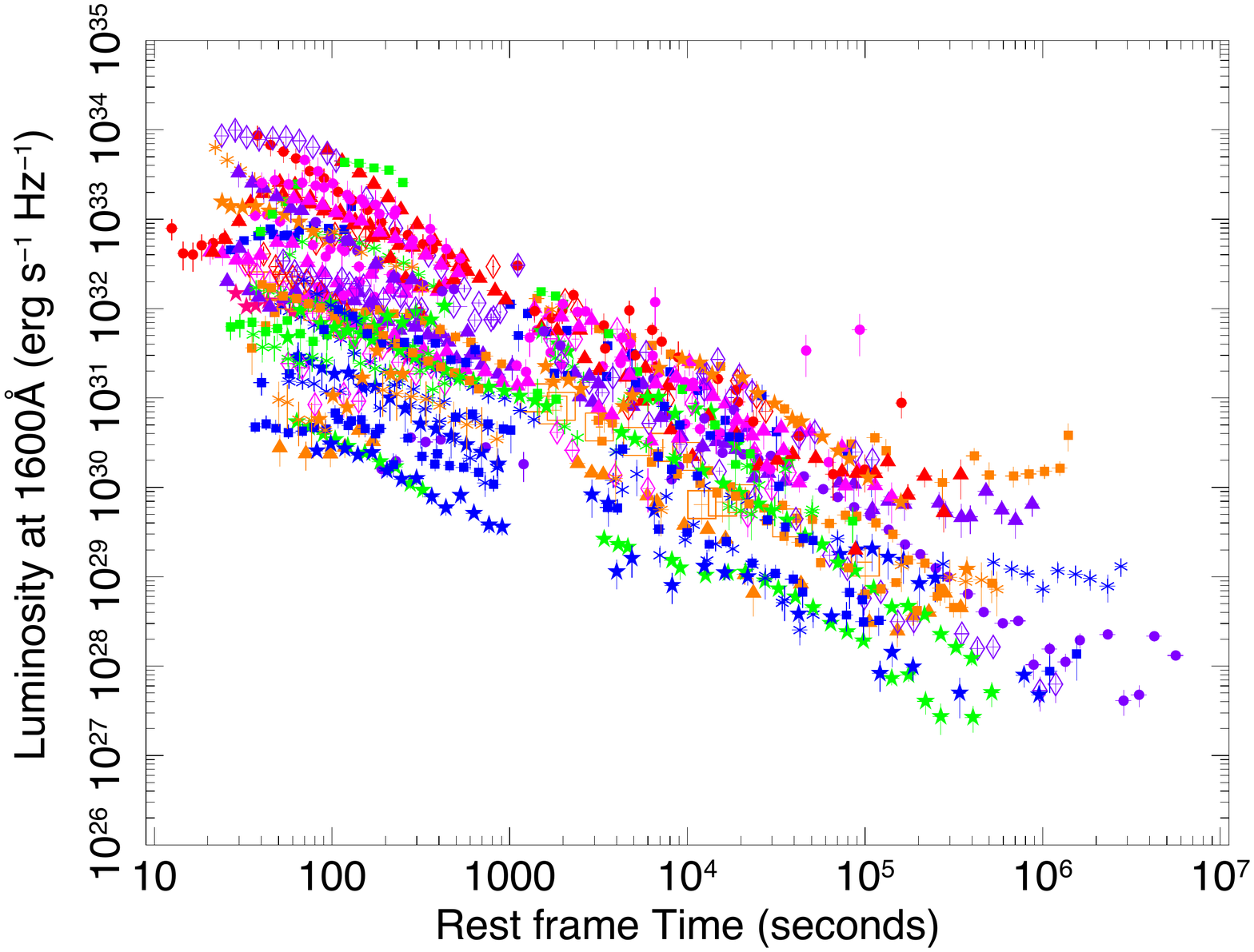}
\includegraphics[width=0.5\textwidth,trim={2cm 14cm 2.5cm 4.cm},clip]{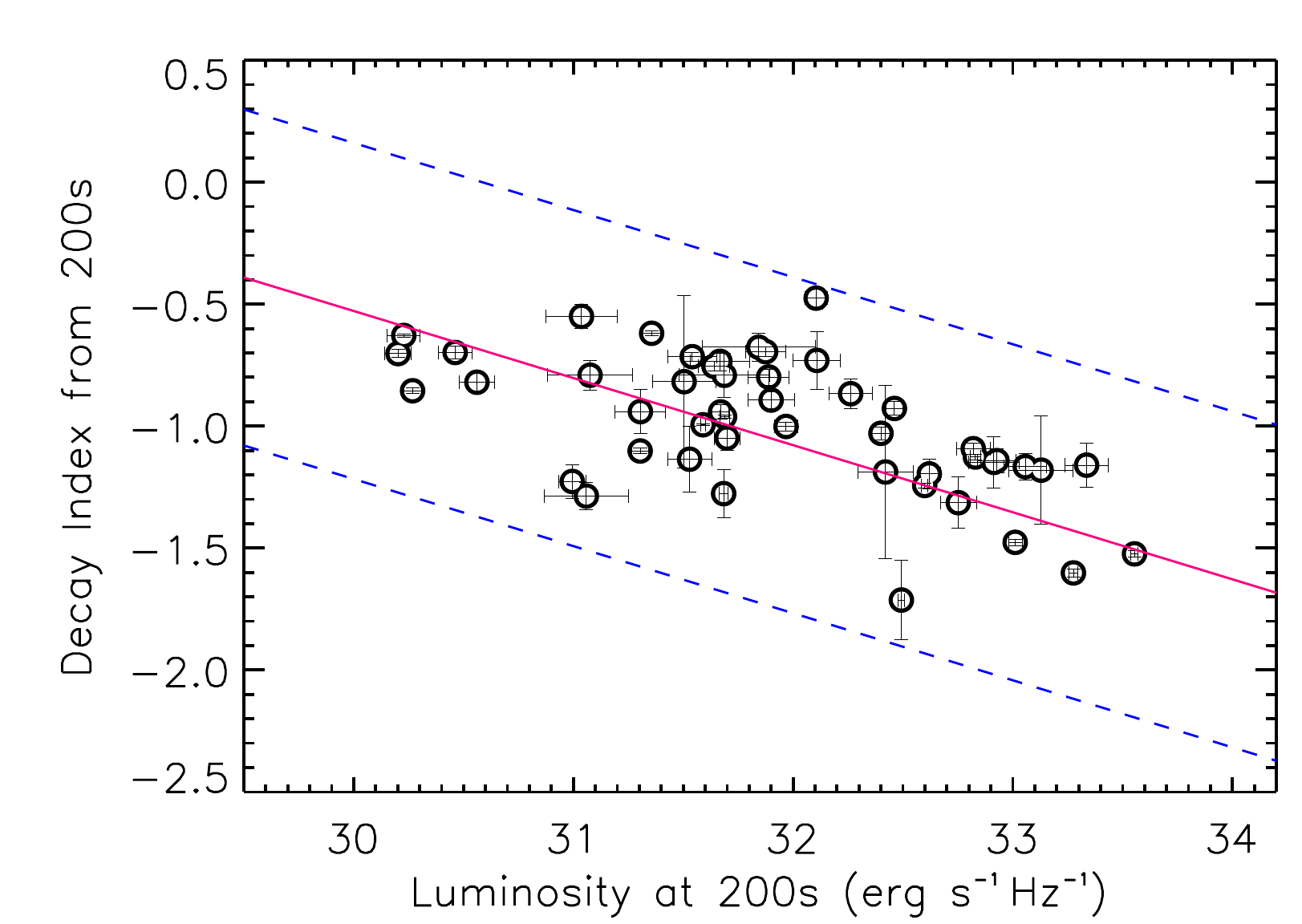}
\caption{Left: Optical luminosity light curves of 56 GRBs at restframe 1600 \AA. Right: Average decay index determined from 48 optical luminosity
light curves after 200 s versus luminosity at 200 s. The red solid line represents the best-fit regression, and the blue dashed line
represents the $3\sigma$ deviation. Figures reproduced from Figures 1 and 2 of \cite{oates12}.} 
\label{fig:UVOTlumlc}
\end{figure}

Using the \swiftu~optical light curves, \cite{oates12} discovered a correlation between the early luminosity (at restframe 200 s) and the average rate of decay (measured with a single power-law from 200 s until the end of observations), see Figure \ref{fig:UVOTlumlc}. This correlation was also shown to be present at X-ray wavelengths \citep{oates15,racusin16}, and most recently, in a small sample of GeV light curves \citep{ryan23}. The early X-ray and optical/UV luminosity also correlates with the isotropic energy of the prompt emission \citep{dav12,mar13,oates15}, implying that the most energetic GRBs have the brightest and fastest decaying afterglows \citep{oates15}. This correlation may be explained via a parameter or mechanism that controls the energy release in GRBs, such that the most energetic GRBs lose their energy quicker, or may be a viewing angle effect, such that GRBs viewed off-axis have fainter and slower decaying light curves \citep{granot02,pan08,oates12}. As this correlation is present in samples of X-ray light curves with and without plateau phases~\citep{racusin16}, it will be important to determine how the luminosity-plateau end time relation relates to the luminosity-decay rate correlation.

Since the luminosity-decay rate and luminosity-plateau end-time correlations relate intrinsic parameters to observed ones, it is hoped that these and other GRB correlations may be used to standardise GRBs, to determine redshifts for GRBs without a spectroscopic or photometrically measured redshift, and for use as standard candles, as was achieved using the Philips relation \citep{phi93} for SNe Ia (e.g., \citep[]{rie22}). This has so far been implemented for several correlations involving prompt emission parameters (e.g., \citep[]{sch03,ama08,cap08,lia08,wan11,izz15,wan15b,muc21}), and the X-ray versions of the luminosity-plateau end-time correlation and the three-parameter 'Fundamental Plane' correlation \cite{car09,car10,dai13,pos14,dai22,wan22,dai23}.

\section{Conclusions/Looking to the Future}
\label{conclusions}
\swift ~has been a highly successful GRB mission for the last 18 years and continues to be the workhorse of the GRB and transient community. The 
flexibility and responsiveness of \swift, in part due to its excellent and dedicated team, ensures that \swift~continues to provide key observations 
to the astronomical community, and continues to be a successful top-ranked NASA mission. 

\swiftu~observations have revolutionised our understanding of the optical/UV emission of GRBs, providing information on the behaviour of the optical/UV emission in the first minutes to hours after the gamma-ray emission, which was unknown prior to \swift. Over the past 18 years, \swiftu~has provided large numbers of optical/UV observations of GRBs with simultaneous X-ray observations. \swiftu~and \swiftx~have highlighted the importance of multi-wavelength observations and analyses, which have been key in unravelling the temporal and spectral behaviour of GRBs. The optical/UV light curves can be more complex than expected pre-\swift,~with optical flares and optical rebrightenings sometimes being superimposed on the expected standard afterglow emission. This behaviour does not always trace the X-ray light curves, and in some cases, it suggests multiple emission components or that a complex jet structure is needed to explain the observations. In particular, 20\% of the early optical light curves are observed initially to rise, which is not observed in the X-ray. 

The number of well-sampled optical/UV light curves observed by \swiftu~has enabled statistical analyses to be performed, resulting in the discovery of correlations between parameters measured from the optical/UV afterglows, and also from the gamma-ray emission, thus connecting to the processes resulting in the prompt and afterglow emission. The optical/UV capabilities of \swiftu~also enable the dust content along the line of sight to be studied, and photometric redshifts to be obtained for GRBs where a spectroscopic redshift has not been possible.

\swiftu~will continue to build the samples of the optical/UV light curves required to explain the collective behaviours of GRBs, and to continue to observe and to detect unique GRBs and transients that push the boundaries of our understanding. There are still many questions to be addressed regarding long GRBs, such as when is the onset of the afterglow for $\sim$80\% of GRBs, what is the structure of the jet, and whether there is any variation in observer viewing angle; \swift~will continue to provide exquisite data to address these issues.

\vspace{6pt} 


\funding{This research received no external funding.}

\dataavailability{The data presented in Section \ref{uvot_overview} of this review are openly available at \url{https://swift.gsfc.nasa.gov/archive/grb\_table/}, accessed 20th October 2022. For all other sections, no new data were created nor analysed, and data sharing is not applicable.} 

\acknowledgments{The author would like to acknowledge and thank all the engineers and scientists that built and calibrated \swiftu, analysed \swiftu~GRB data, and supported and planned \swiftu~observations. This review paper would not be possible without their time and dedication. This research has made use of data obtained from the High Energy Astrophysics Science Archive Research Center (HEASARC) and the Leicester Database and Archive Service (LEDAS), provided by NASA's Goddard Space Flight Center and the School of Physics and Astronomy, University of Leicester, UK, respectively.}

\conflictsofinterest{The authors declare no conflicts of interest.} 

\clearpage 
\abbreviations{Abbreviations}{
The following abbreviations are used in this manuscript:\\

\noindent 
\begin{tabular}{@{}ll}
GRB & Gamma-ray burst \\
IR   & Infrared \\
LGRB & Long gamma-ray burst \\
SGRB & Short gamma-ray burst \\
SN(e)  & Supernova(e) \\ 
UV   & Ultra-violet \\
UVOT & Ultra-Violet/Optical Telescope \\  
XRT  & X-Ray Telescope \\

\end{tabular}
}

\begin{adjustwidth}{-\extralength}{0cm}

\reftitle{References}

\bibliography{lGRBs_18yearUVOT}

\end{adjustwidth}
\end{document}